\documentclass[prd,tightenlines,nofootinbib,superscriptaddress]{revtex4}

\usepackage{slashbox}
\usepackage{array}
\usepackage{amsfonts,amssymb,amsthm,bbm}

\usepackage{amsmath}

\usepackage{hyperref}

\usepackage{subcaption}
\usepackage{tikz}
\usetikzlibrary{calc}
\usetikzlibrary{decorations.pathmorphing}
\usetikzlibrary{shapes.geometric}
\usetikzlibrary{arrows,decorations.markings}

\usepackage{color,psfrag}

\newcommand{\C}{{\mathbb C}}
\newcommand{\N}{{\mathbb N}}
\newcommand{\R}{{\mathbb R}}

\newcommand{\cL}{{\mathcal L}}

\newcommand{\cM}{{\mathcal M}}

\newcommand{\cR}{{\mathcal R}}

\newcommand{\cP}{{\mathcal P}}

\newcommand{\cV}{{\mathcal V}}
\newcommand{\cD}{{\mathcal D}}
\newcommand{\cC}{{\mathcal C}}

\newcommand{\SU}{\mathrm{SU}}

\newcommand{\SL}{\mathrm{SL}}

\newcommand{\U}{\mathrm{U}}

\newcommand{\be}{\begin{equation}}
\newcommand{\ee}{\end{equation}}
\newcommand{\beq}{\begin{eqnarray}}
\newcommand{\eeq}{\end{eqnarray}}
\newcommand{\bes}{\begin{eqnarray}}
\newcommand{\ees}{\end{eqnarray}}

\newcommand{\mat} [2] {\left ( \begin{array}{#1}#2\end{array} \right ) }

\newcommand{\su}{{\mathfrak{su}}}

\renewcommand{\sl}{{\mathfrak{sl}}}

\newcommand{\la}{\langle}
\newcommand{\ra}{\rangle}

\newcommand{\tr}{{\mathrm{Tr}}}
\newcommand{\f}{\frac}

\def\nn{\nonumber}
\def\pp{\partial}

\def\vphi{\varphi}
\def\eps{\epsilon}

\newcommand{\id}{\mathbb{I}}

\def\act{\triangleright}

\def\vX{\vec{X}}
\def\vsigma{\vec{\sigma}}
\def\arr{\rightarrow}
\def\act{\,\triangleright\,}
\def\bz{\bar{z}}
\def\tX{\tilde{X}}
\def\vcC{\vec{\cC}}
\def\tw{\tilde{w}}

\def\tm{\tilde{m}}
\def\vK{\vec{K}}

\def\vJ{\vec{J}}

\def\up{\uparrow}
\def\down{\downarrow}

\newcommand{\alink}[4]
{\draw[decoration={markings,mark=at position 0.6 with {\arrow[scale=1.5,>=stealth]{>}}},postaction={decorate}] (#1) -- node[#3,pos=.5]{$#4$}(#2)}
\newcommand{\link}[2]
{\draw[decoration={markings,mark=at position 0.6 with {\arrow[scale=1.5,>=stealth]{>}}},postaction={decorate}] (#1) --(#2)}

\def\centerarc[#1](#2)(#3:#4:#5)% Syntax: [draw options] (center) (initial angle:final angle:radius)
    { \draw[#1] ($(#2)+({#5*cos(#3)},{#5*sin(#3)})$) arc (#3:#4:#5); }
    
\def\centerarcnodes[#1](#2)(#3:#4:#5)(#6,#7)% Syntax: [draw options] (center) (initial angle:final angle:radius)
    {\coordinate(#6) at ($(#2)+({#5*cos(#3)},{#5*sin(#3)})$);
    \coordinate(#7) at ($(#2)+({#5*cos(#4)},{#5*sin(#4)})$);
    \draw[#1] ($(#2)+({#5*cos(#3)},{#5*sin(#3)})$) arc (#3:#4:#5); }

\def\angcircle(#1)(#2)(#3:#4) {\coordinate(#1) at ($(#2)+({#4*cos(#3)},{#4*sin(#3)})$); }

\begin{document}

\title{Area Propagator \& Boosted Spin Networks in Loop Quantum Gravity}

\author{{\bf Etera R. Livine}}\email{etera.livine@ens-lyon.fr}
\affiliation{Perimeter Institute for Theoretical Physics, 31 Caroline Street North, Waterloo, Ontario, Canada N2L 2Y5}
\affiliation{Universit\'e de Lyon, ENS de Lyon,  Laboratoire de Physique, CNRS UMR 5672, F-69342 Lyon, France}

\date{\today}

\begin{abstract}

Quantum states of geometry in loop quantum gravity are defined as spin networks, which are graph dressed with $\SU(2)$ representations. A spin network edge carries a half-integer spin, representing basic quanta of area, and the standard framework imposes an area matching constraint along the edge: it carries the same spin at its source and target vertices. In the context of coarse-graining, or equivalently of the definition of spin networks as projective limits of graphs, it appears natural to introduce excitations of  curvature along the edges.
An edge is then treated similarly to a propagator living on the links of Feynman diagrams in quantum field theory: curvature excitations create little loops -tadpoles- which renormalize it. This relaxes the area matching condition, with different spins at both ends of the edge. We show that this is equivalent to combining  the usual $\SU(2)$ holonomy along the edge with a Lorentz boost into $\SL(2,\C)$ group elements living on the spin network edges, underlining the fact that the Ashtekar-Barbero connection carries extrinsic curvature degrees of freedom. This finally leads us to introduce a new notion of area waves in loop quantum gravity.

\end{abstract}

\maketitle
%%%%%%%%%%%%%%%%%%%%%%%%%%%%%%%%%%%%%%%%%%%%%%%%%%%%%%%%
%\tableofcontents

%%%%%%%%
%\section*{Introduction}
%%%%%%%%

A key issue of the Loop Quantum Gravity approach to quantum gravity is the coarse-graining of the theory. It is a crucial unvaoidable step in order to understand its continuum limit towards general relativity. And it is necessary to open the door to go further and understand the phenomenology and physics of the theory, derive the renormalization flow of the theory, explore its phase diagram and the new possible phases of geometry and matter that it could lead to and classify the universality classes of microscopic quantum  dynamics of the geometry.

Quantum states of geometry in Loop quantum gravity are spin networks. Algebraically, they are defined as graph dressed with $\SU(2)$ irreducible representations and intertwiners from the theory of spin recoupling \cite{Rovelli:1994ge,Rovelli:1995ac}. Geometrically,  they are understood as the quantized version of discrete twisted geometries \cite{Freidel:2010aq}, which generalize Regge triangulations \cite{Dittrich:2012rj,Dupuis:2012yw,Freidel:2013bfa,Freidel:2018pvm}. Physically, they can be interpreted as networks of space points, related by links indicating possible flows of information, thereby providing a cartography of all possible processes that could occur in the  quantum 3d space.

While spin networks are supposed to provide an accurate picture of the quantum geometry at the Planck scale, numerous works shows that their coarse-graining naturally leads to generalizing their mathematical and physical definitions, for instance to loopy spin networks \cite{Charles:2016xwc}, $q$-deformed spin networks \cite{Dupuis:2013lka,Dupuis:2014fya,Charles:2016xzi}, networks dressed with representations of the Drinfeld double $\cD\SU(2)$ \cite{Dittrich:2014wpa,Bahr:2015bra,Delcamp:2016yix} or conformal blocks \cite{Markopoulou:1997hu,Freidel:2009nu,Freidel:2016bxd}.
The main process is that the connection curvature builds up during the coarse-graining procedure: initially located around the loops of the graph, coarse-graining should incorporate it, step after step, into the basic elements of the spin networks, its nodes and links. This should lead to dressed vertices and edges for spin networks, the same way that Feynman diagram vertices and links gets renormalized in perturbative quantum field theory.

Up to now, work on the coarse-graining of spin networks has focused on vertices. Indeed, vertices represent at the fundamental level the quanta of 3d volume, i.e. the elementary pieces of 3d space, and they then  represent at the effective level finite bounded 3d region which have been coarse-grained to a single point. More technically, a bare vertex carries a closure constraint, which allows it to be interpreted geometrically as dual to a convex polyhedron living in the 3d flat Euclidean space \cite{Barbieri:1997ks,Freidel:2010tt,Bianchi:2010gc,Livine:2013tsa}. It defines a fundamental piece of 3d volume. Now, let us consider a finite region of space, defined as a bounded subset of a spin network, and let us coarse-grain to a single vertex assuming that we do not have a high enough resolution to probe the deep internal structure of that region. It was shown in \cite{Livine:2006xk,Charles:2016xwc} that, after a gauge-fixing procedure used as the first step of coarse-graining, the region can be pictured as a vertex with little loops attached to it and which represent the curvature excitations that developed around the loop of the coarse-grained region. Such vertex with an arbitrary number of little loops defines a dressed spin network vertex.  From the point of view of the boundary edges linking the dressed vertex to the other vertices outside the region and tracing over the little loops, it appears that the closure constraint is relaxed and we lose a priori the straightforward geometrical interpretation in terms of dual flat polyhedra \cite{Charles:2016xwc}. It is nevertheless possible to identify a unique Lorentz boost to a frame in which the geometrical fluxes close once again: in this boosted frame, we recover the closure constraint and the interpretation of the dressed vertex as dual to a polyhedron \cite{Freidel:2010tt,Livine:2013gna}.

\medskip

The present work turns to the fate of spin network links. At the classical level, the link carries the $\SU(2)$ holonomy of the Ashtekar-Barbero connection, which defines the change of frame or 3d transport between the link's source vertex and its target vertex. Upon quantization, the link acquires an extra-label, a half-integer spin, which defines the wave mode of the $\SU(2)$ holonomy and more importantly is understood as the quanta of area \footnotemark{} carried by the link \cite{Rovelli:1994ge,Ashtekar:1996eg}.
\footnotetext{
It is geometrically understood as the area of a surface transverse to the link, i.e. the 2d interface between the two blocks of 3d volume dual to the source and target vertices of the link, see e.g. \cite{Freidel:2018pvm} for a recent detailed discussion.
}
Here, we stress that any point along the link can be thought as a bivalent vertex (due to the cylindrical consistency of the spin network functional e.g. \cite{Ashtekar:1994mh}). At the effective level during the coarse-graining, we can thus  attach to any point along a link  little loops, which carry curvature excitations. This leads to a new notion of {\it dressed spin network edges}. The main effect is that the spin at one end of the edge does not need to match the spin at the other end and we can obtain a non-trivial propagation of the spin along the edge from its source to its target. Since the spin gives the area at the quantum level, we dub this the area propagator. Just like the renormalization of the propagator by Feynman diagrams in quantum field theory, this area propagator should carry crucial information about the coarse-graining flow and renormalization of spin network states.

We further show, at both classical and quantum levels, that the insertion of little loops along the edges can be mathematically interpreted as the spin network edges now carrying a $\SL(2,\C)$ holonomy instead of simply a $\SU(2)$ holonomy.
This hints towards the necessity to rethink loop quantum gravity's $\SU(2)$ spin networks  more in terms of Lorentz connection, as in the spinfoam framework \cite{Engle:2007wy,Ding:2010ye,Dupuis:2010jn}. This underlines the embedding-dependence of the Ashtekar-Barbero connection and holonomies, with the crucial role of the extrinsic curvature \cite{Samuel:2000ue,Alexandrov:2001wt,Geiller:2011cv,Geiller:2012dd,Charles:2015rda}. This important interplay between intrinsic and extrinsic geometry in loop quantum gravity means that coarse-graining spin networks is not only about coarse-graining the intrinsic geometry of the space manifold but it should involve coarse-graining the embedding of the canonical spatial hypersurface in the surrounding space-time. At the technical level, this translates into the fact that non-trivial Ashtekar-Barbero  holonomies can be thought alternatively as intrinsic geometry or extrinsic curvature or both, which explains that we can compensate defects of the intrinsic geometry  appearing while coarse-graining spin networks by Lorentz boosts which change the local embedding of the spin network in the 3+1-d geometry. 
%Choosing one or the other probably affect coarse-graining procedure and geometric interpretation.

\medskip

The first section of this short paper deals with the classical counterpart of spin networks, that is how to dress and coarse-grain links of twisted geometries. We explain how the insertion of loops along an edge accounts for a non-trivial curvature excitation, how it leads to relaxing the area-matching constraint along the edge and can be represented in terms of $\SL(2,\C)$ holonomies. These Lorentz group elements extend and upgrade the $\SU(2)$ holonomies to the case when the area-matching constraint is not enforced anymore.

The second section deals with the quantum case. The insertion of loops along a spin network edge allows to change the spin along the edge, which is the quantum counterpart of relaxing the area-matching constraint. We show that a loop insertion, translated in terms of Clebsch-Gordan coefficients, is equivalent to the insertion of a $\sl(2,\C)$ boost generator. We further analyze the action of a $\SL(2,\C)$ group element on a spin network edge and the resulting probability distribution for the spin at the target vertex in terms of the spin at the source vertex. This leads to new notions of spin diffusion and spin wave along spin network edges.

%%%%%%%%
\section{Coarse-graining Twisted Geometries}
\label{classical}
%%%%%%%%

%%%
\subsection{Twisted geometries in a nutshell}
%%%

Let us start with the definition of the phase space of twisted geometries \cite{Freidel:2010aq,Freidel:2010bw,Borja:2010rc}. They describe the discrete geometries underlying the spin network states of quantum geometry in loop quantum gravity. They are mathematically defined as networks of $\SU(2)$ group elements dressed with compatible geometrical data.

More precisely, we consider a (closed) oriented graph $\Gamma$ and introduce the data of a complex 2-vector, or spinor, $z^v_{e}\in\C^2$ on each half-edge, i.e. on at the end of every link or edge $e$ attached to a node or vertex $v$:
\be
|z^v_{e}\ra=\mat{c}{(z^v_{e})_{0}\\(z^v_{e})_{1}}\in\C^2
\,,\quad
\la z^v_{e}|=\mat{cc}{(\bz^v_{e})_{0}&(\bz^v_{e})_{1}}
\,.
\ee
So each oriented edge $e$ carries two such spinors, respectively attached to its source and target vertices, which we write $z^s_{e}$ and $z^t_{e}$, where we write  $s$ for the source vertex $v=s(e)$ and $t$ for the target vertex $v=t(e)$.
Each spinor variable is endowed with the canonical Poisson bracket:
\be
\{z_{A},\bz_{B}\}=-i\delta_{AB}
\,,\quad
\{z_{A},z_{B}\}=\{\bz_{A},\bz_{B}\}=0
\,,
\ee
with $A,B\in\{0,1\}$ and where we have dropped the indices $e$ and $v$ for simplify the notations.

The spinors satisfy constraints reflecting the combinatorics of the graph $\Gamma$. There are two sets of constraints, corresponding to the graph edges and vertices:
\begin{itemize}
\item Matching constraints along the edges $e$:
\be
\cM_{e}\equiv\la z^s_{e} | z^s_{e}\ra-\la z^t_{e} | z^t_{e}\ra=0
\ee

\item Closure constraints at the vertices $v$:
\be
\sum_{e\ni v} |z^v_{e}\ra\la z^v_{e}| \propto \id
\quad\textrm{or equivalently}\quad
\vcC_{v}\equiv\sum_{e\ni v} \la z^v_{e}| \vsigma|z^v_{e}\ra=0
\,,
\ee
\end{itemize}
where $\vsigma$ is a 3-vector notation for the  three Pauli matrices $\sigma_{a}$ normalized so that they square to the identity.

These constraints are all first class. The matching constraints generate $\U(1)$ gauge transformations on each edge $e$:
\be
| z^s_{e}\ra\mapsto e^{i\theta_{e}} | z^s_{e}\ra
\,,\quad
| z^t_{e}\ra\mapsto e^{-i\theta_{e}} | z^t_{e}\ra
\,,
\ee
where $e^{i\theta_{e}}\in\U(1)$ is an arbitrary phase, while the closure constraints generate $\SU(2)$ gauge transformations around every vertex $v$:
\be
|z^v_{e}\ra\mapsto\, h_{v}|z^v_{e}\ra
\,,
\ee
where $h_{v}\in\SU(2)$ is an arbitrary  group element acting in the fundamental $\SU(2)$ representation as 2$\times$2 matrices.

In order to recover the usual holonomy-flux phase space of loop quantum gravity variables on the graph $\Gamma$, one solves the matching constraints and define  $\U(1)$-invariant observables (i.e. which Poisson-commute with the matching constraints). We introduce 3-vectors:
\be
\vX^v_{e}=\la z^v_{e}|\vsigma|z^v_{e}\ra \in\R^3
\,,\qquad
\{\cM_{e},\vX^{s,t}_{e}\}=0
\,,
\ee
and $\SU(2)$ group elements along the edges:
\be
g_{e}
=
\f{|z^t_{e}{]}\la z^s_{e}|-|z^t_{e}\ra {[}z^s_{e}|}{\sqrt{\la z^s_{e} | z^s_{e}\ra\,\la z^t_{e} | z^t_{e}\ra}}
\in\SU(2)
\,,\qquad
\{\cM_{e},g_{e}\}=0
\,,
\ee
where we have introduced a dual spinor notation:
\be
|z]=\mat{c}{\bz_{1}\\ -\bz_{0}}
\,,\quad
[z|=\mat{cc}{z_{1} & -z^0}
\,.
\ee
These are the usual holonomy-flux variables of loop quantum gravity. The 3-vectors $\vX^v_{e}$ define the geometrical fluxes (or discretized triad) around every vertices $v$, while the $\SU(2)$ group elements $g_{e}$ are the  holonomies of the Ashtekar-Barbero connection integrated along the edges $e$, which give the transport from one node to another. The geometrical and transport data are compatible in the sense that the the group element $g_{e}$ maps the source flux $\vX^s_{e}$ onto the target flux $\vX^t_{e}$ up to a switch of orientation:
\be
\left|\begin{array}{l}
g_{e}|z^s_{e}\ra=|z^t_{e}] \vspace*{1mm}\\
g_{e}|z^s_{e}]=-|z^t_{e}\ra
\end{array}\right.
\,,\qquad
g_{e} X^s_{e} g_{e}^{-1}=-X^t_{e}
\,,
\ee
where we have repackaged the 3-vectors $\vX$ as 2$\times$2 traceless Hermitian matrices $X=\vX\cdot\vsigma$:
\be
\forall z\in\C^2\,,\quad
X(z)=\vX(z)\cdot\vsigma
=2|z\ra \la z|-\la z|z\ra\id
%=|z\ra \la z|-\f12\la z|z\ra\id
\,,\quad
\tr\,X=0
\,,\quad
\vX=\f12\tr\,\vsigma X
\,.
\ee
In particular, the source and target fluxes have equal norm, $| X^s_{e}|=|X^t_{e}|$, which reflects the matching constraints.

The Poisson brackets of the holonomy-flux variables form a $T^*\SU(2)$ algebra on each edge:
\be
\{g_{e},g_{e'}\}=0
\,,\quad
\{(X_{e}^{v})_{a},(X_{e}^{v})_{b}\}=2\eps^{abc}(X_{e}^{v})_{c}
\,,\quad
\{\vX^s_{e},g_{e}\}=-i g_{e}\vsigma
\,,\quad
\{\vX^t_{e},g_{e}\}=+i \vsigma g_{e}
\ee
while the closure constraints now read
\be
\forall v\,,\,\,
\sum_{e\ni v}\vX^v_{e}=0
\ee
and generate $\SU(2)$ gauge-transformations on the fluxes and holonomies:
\be
X^v_{e}\mapsto h_{v} X^v_{e} (h_{v})^{-1}
\,\quad
g_{e}\mapsto h_{t(e)} g_{e} h_{s(e)}^{-1}
\,.
\ee
These networks of holonomies and fluxes are usually interpreted as discrete geometries in the framework of twisted geometries. Locally, the closure constraint around a vertex $v$ implies the existence of a unique convex polyhedron such that the fluxes $\vX^v_{e}$ are the normal vectors to the polyhedron's faces \cite{Bianchi:2010gc}. These polyhedra are the basic building blocks of the discrete geometry. They are glued together to form the 3d space geometry. Then the $\SU(2)$ group elements $g_{e}$ describe the change of 3d frame from one polyhedron to a neighboring one. 
This twisted geometry picture provide a generalization of Regge geometries that account for the possible non-trivial torsion of the Ashtekar-Barbero connection, which encodes data on the extrinsic curvature of the 3d manifold. This is taken into account by the twist angle on each edge, which is one component of the $\SU(2)$ group elements, and the apparent mismatch of the shape of the polyhedra faces across the network edges \cite{Freidel:2010aq,Dittrich:2012rj,Haggard:2012pm}.
This twisted geometry picture can be refined in the spinning geometry framework \cite{Freidel:2013bfa} and has recently been enhanced with more geometric data in the context of bubble networks \cite{Freidel:2018pvm}.

%%%
\subsection{Dressing links with curvature}
%%%

An important topic of research in loop quantum gravity is the coarse-graining of twisted geometries and their quantum counterpart - the spin network states. This is a crucial point towards establishing a rigorous continuum limit (in which we should recover general relativity or a suitable modification) and renormalization for quantum states of geometry and their dynamics.

\medskip

Previous works \cite{Livine:2013gna,Charles:2016xwc} focused on the vertices or nodes of twisted geometries and underlined the necessity to extend their algebraic structure to be able to account for the curvature and torsion potentially building up during the coarse-graining process. Indeed, every node of a twisted geometry is interpreted as carrying a flat polyhedron (embedded in $\R^3$) reconstructed from the fluxes $\vX^e_{v}$ meeting at that vertex and satisfying the closure constraint $\sum_{e\ni v}\vX^e_{v}=0$. In this context, curvature emerges as non-trivial $\SU(2)$ holonomies around loops: the $\SU(2)$ group elements $g_{e}$ define the transport from one polyhedron to the next and their product $\prod_{e\in\cL} g_{e}$ around a closed path $\cL$ of polyhedra is a priori arbitrary and can be non-trivial.

\begin{figure}
%\centering
\includegraphics[width=60mm]{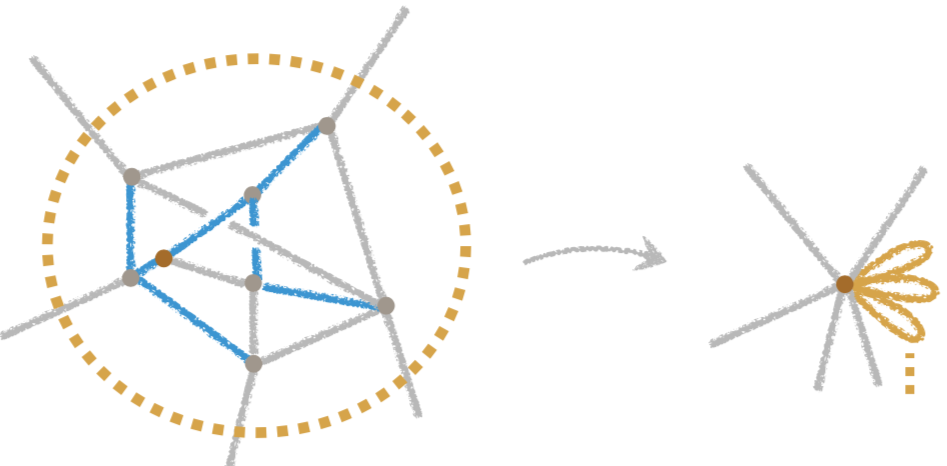}
\caption{Coarse-graining a bounded region of a spin network into a single dressed vertex: following the procedure described in \cite{Freidel:2002xb,Livine:2006xk,Livine:2013gna,Charles:2016xwc}, one gauge-fixes the $\SU(2)$ holonomies along a maximal tree (in blue) of the bulk subgraph within the to-be-coarse-grained region, which reduces the region to a single vertex with little loops (in orange) attached to it corresponding to the non-gauge-fixed links. These little loops carry local curvature excitations at that vertex. Coarse-graining then amounts to deciding how much combinatorial data we retain from the original bulk graph and how much algebraic data we retain from the $\SU(2)$ holonomies living on the little loops \cite{Charles:2016xwc,Feller:2017ejs}. }
\label{fig:coarsegrainedvertex}
\end{figure}

Then let us consider the coarse-graining of a bounded finite region $\cR$ of a twisted geometry to a single renormalized vertex $v$, as illustrated on figure \ref{fig:coarsegrainedvertex}. As explained in \cite{Livine:2013gna,Charles:2016xwc}, curvature builds up around the non-trivial loop of the graph within the to-be-coarse-grained region, so that the region is coarse-grained to a single vertex to which are attached both the boundary edges $e\in\pp\cR$ and ``little loops'' $\ell$ representing those non-trivial loops carrying curvature. 

The primary effect of those little loops is to imply a defect in the closure constraint, such that now $\sum_{e\in\pp\cR} X_{e}$ does not vanish anymore. As shown in \cite{Freidel:2010tt,Livine:2013tsa,Livine:2013gna}, one can nevertheless perform a unique  boost $\Lambda_{v} \in\SL(2,\C)$ to map a non-closed flux configuration  $\sum_{e} X_{e}\ne 0$ to a closed configuration $\sum_{e}\tX_{e}=0$ with $\tX_{e}=\Lambda_{v}\triangleright X_{e}$. This can be interpreted as a local change of slicing, re-absorbing the local curvature of the Ashtekar-Barbero connection into the extrinsic curvature of the 3d spatial slice. 

This previous work showed the necessity of introducing dressed vertices, extending the original vertices of twisted geometry in extra algebraic data so as to account for local curvature excitations at the vertices.

\medskip

The goal of the present work is to apply the same line of thought to the edges of twisted geometries and understand how they should be generalized in the context of coarse-graining twisted geometries.

\begin{figure}
%\centering
\includegraphics[width=60mm]{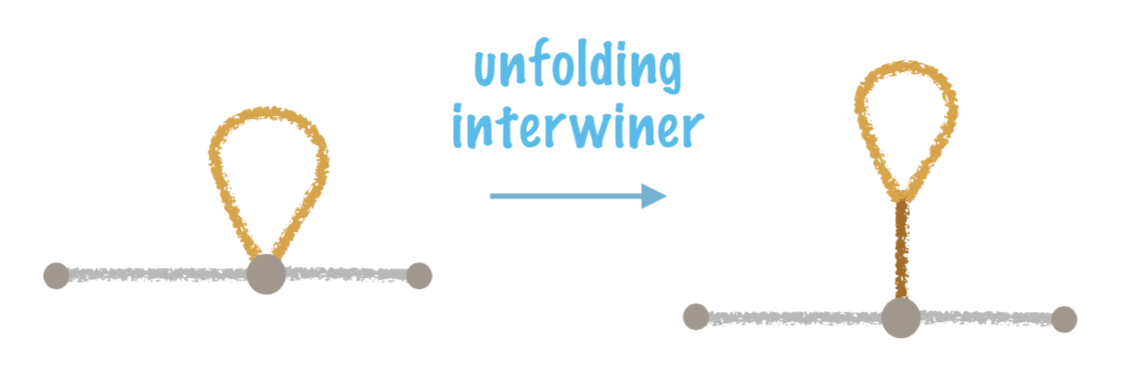}
\caption{Any point on a spin network edge can be considered as a bivalent vertex (by cylindrical consistency), which can develop local curvature excitation materialized as loops: we can unfold the intertwiner attaching the loop to the edge in the spin basis with a virtual edge linking the loop to the edge.}
\label{fig:looponedge}
\end{figure}

\begin{figure}[h!]
%\centering
\begin{tikzpicture}[scale=1.2]

%\arrow[scale=1.5,>=stealth]{>}}
%\draw [->,>=stealth]  (0,0)node {$\bullet$}node[below]{$|z^s\ra$}--(1,0) ;
\link{0,0}{2,0};
\draw (0,0)node {$\bullet$}node[below]{$|z^s\ra$};
\draw  (0.9,0.05) node[above]{$g^s$};
%\draw  (1,0)--(2,0) ;
\draw  (1.65,-0.25) node{$|w^s]$};

\link{2,0}{4,0};
\draw  (2.3,-0.23) node{$|w^t\ra$};
\draw  (3.2,0.05)node[above]{$g^t$};
\draw  (4,0) node {$\bullet$}node[below]{$|z^t]$} ;

\coordinate(A) at (2,0);
\draw (A) node {$\bullet$};
\draw[scale=4,decoration={markings,mark=at position 0.51 with {\arrow[scale=1.5,>=stealth]{>}}},postaction={decorate}] (A) to[in=+45,out=+135,loop]  node[midway,above]{$h$}(A) ;
\draw (A)++(-0.5,0.25) node{$|\alpha\ra$};
\draw (A)++(+0.5,0.25) node{$|\beta]$};

\end{tikzpicture}

\caption{Insertion of a loop carrying the $\SU(2)$ holonomy $h$ in the middle of a twisted geometry edge: the closure relation between the four spinors $\alpha,\beta,w^s,w^t$ allows to relax the area-matching constraint along the edge which initially enforced that the source and target spinors, $z^s$ and $z^t$, have equal norms.} 
\label{loopyedge}

\end{figure}
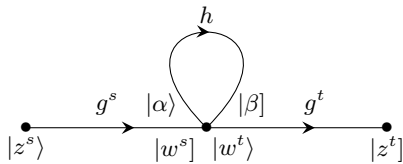

Initially, at the fundamental level in the original definition given in the previous section, an edge $e$ carries a  group element $g_{e}\in\SU(2)$ mapping the spinor at its source to the spinor at its target, $g_{e}\,|z^s_{e}\ra=|z^t_{e}]$. Let us now imagine possible curvature excitations along the edges, which would develop as little loops or tadpoles attached to the edge\footnotemark{} as illustrated on figure  \ref{fig:looponedge}.
\footnotetext{
This is natural from the point of view of the cylindrical consistency imposed in loop quantum gravity: an edge is equivalent to a sequence of bivalent vertices. Since vertices can acquire  self-loops under coarse-graining, so should the edges.
}
Considering a single tadpole excitation as on figure \ref{loopyedge}, it creates a defect such that the source spinor $z^s_{e}$ and target spinor $z^t_{e}$ can not be simply related by a $\SU(2)$ group element anymore due to the insertion of  the little loop along the edge. Let us look in detail into these structures.

%%%
\subsection{From $\SU(2)$ to $\SL(2,\C)$ holonomies}
%%%

As shown on figure \ref{loopyedge}, the propagation of the spinor from the source to the target of the edge is affected by the little loop insertion. Instead of the straightforward $g\,|z^s\ra=|z^t]$ with $g\in\SU(2)$ we have a sequence of three steps: two half-edges carrying an usual $\SU(2)$ holonomy glued together with a defect due to the little loop.
More precisely, we have the following transport relations:
\be
\left\{
\begin{array}{llcl}
g^s & |z^s\ra &=&  |w^s] \\
h & |\alpha\,\ra &=&  |\beta\,] \\
g^t & |w^t\ra &=&  |z^t] \\
\end{array}
\right.
\ee
combined with the closure constraint attaching the little loop to the edge:
\be
\label{closure}
\cC=X(\alpha)+X(\beta)+X(w^s)+X(w^t)=0
\,.
\ee

In the special case when the little loop is trivial, i.e. if the $\SU(2)$ holonomy carried to the loop is trivial, $h=\id$, then the transport condition $|\alpha\ra =  |\beta]$ implies that $X(\alpha)+X(\beta)=0$. So the little loop effectively decouples from the edge and the closure constraint along the edge reduces to $X(w^s)+X(w^t)=0$. This means that we have a straightforward bivalent node in the middle of the edge, which is interpreted as pure gauge. More precisely, the condition  $X(w^s)+X(w^t)=0$ along the edge implies that the two spinors $|w^s]$ and $ |w^t\ra$ are equal up to a phase. In particular, they have equal norm, $\la w^s|w^s\ra=\la w^t|w^t\ra$, and they can be related by a $\SU(2)$ transformation:
\be
%\left\{
\left|
\begin{array}{lll}
|w^t\ra
&=
e^{i\vphi}|w^s]
&=
g_{w}|w^s]
\vspace*{1mm}\\
|w^t]
&=
-e^{-i\vphi}|w^s\ra
&=
g_{w}|w^s\ra
\end{array}\right.
\qquad\textrm{with}\quad
g_{w}
=
\f{|w^t\ra [ w^s|-|w^t]\la w^s|}{\sqrt{\la w^s|w^s\ra\la w^t|w^t\ra}}
=
\f{e^{i\vphi}|w^s][w^s|+e^{-i\vphi}|w^s\ra\la w^s|}{\la w^s|w^s\ra}
\,\, \in
 \SU(2)
\ee
where we wrote the phase shift as the action of a $\SU(2)$ group element. So overall the edge is carrying the $\SU(2)$ group element $g=g^tg_{w}g^s$, which maps the source spinor $|z^s\ra$ onto the target spinor $|z^t]$ as expected:
\be
\left|\begin{array}{lllll}
g|z^s\ra &
= g^tg_{w}g^s |z^s\ra &
= g^tg_{w}|w^s] &
= g^t |w^t\ra &
= |z^t]
\vspace*{1mm}\\
g|z^s] &
= g^tg_{w}g^s |z^s] &
= -g^tg_{w}|w^s\ra &
= g^t |w^t] &
= -|z^t\ra
\end{array}\right.
\ee

\medskip

However, in the general case, when the little loop carries an arbitrary holonomy $h$, the spinors $\alpha$ and $\beta$ do not decouple from the edge and the closure constraint attaching the loop to the edge induces a non-trivial propagation from $w^s$ to $w^t$. The fact that $X(w^s)+X(w^t)=-\big{[}X(\alpha)+X(\beta)\big{]}\ne 0$ does not a priori vanish anymore means that, first, the two spinors $w^s$ and $w^t$ do not have equal norm, $\la w^s|w^s \ra\ne \la w^t|w^t \ra$, and thus can not be related by a $\SU(2)$ transformation.
In particular, this means that the norm-matching condition on the coarse-grained edge linking $z^s$ to $z^t$ is violated, $\la z^s|z^s \ra- \la z^t|z^t \ra\ne 0$.

Following the method introduced for intertwiners of arbitrary valence \cite{Freidel:2010tt,Livine:2013tsa}, we can start from the non-closed 2-valent configuration, defined by the two spinors $\{|w^s\ra,|w^t\ra\}$ with a closure defect $X(w^s)+X(w^t)\ne 0$, and there exists a unique pure boost\footnotemark{} $B\in\SL(2,\C)$, $B=B^\dagger$, such that the new configuration $\{B|w^s\ra,B|w^t\ra\}$ satisfies the 2-valent closure relation $X(B\act w^s)+X(B\act w^t)= 0$.
\footnotetext{
More precisely, the 2$\times$2 matrix $|w^s\ra\la w^s|+|w^t\ra\la w^t|$ is Hermitian and  positive (as long as the two spinors do not vanish), so its square-root is well-defined (e.g. by diagonalizing it):
\be
|w^s\ra\la w^s|+|w^t\ra\la w^t|
\,=\,
\lambda\, M^2
\qquad\textrm{with}\quad
\lambda >0
\,,\,\,
M=M^\dagger
\,,\,\,
\det\,M=1
\,,\,\,
M\in\SL(2,\C)
\,.
\ee
$M$ and its inverse $B\equiv M^{-1}$ are pure boosts in $\SL(2,\C)$ and be used to map the spinors onto a closed configuration:
\be
|\tw^{s,t}\ra=B\,|w^{s,t}\ra
\,,\quad
|\tw^s\ra\la \tw^s|+|\tw^t\ra\la \tw^t|
=
B\,\big{(}
|w^s\ra\la w^s|+|w^t\ra\la w^t|
\big{)}\,
B^\dagger
=
\lambda\,\id
\qquad\textrm{thus}\quad
X(\tw^s)+X(\tw^t)= 0
\,.
\ee
}
Once we are back in the special case of a trivial 2-valent vertex, we know that the two boosted spinors $B|w^s\ra$ and $B|w^t\ra$ are equal up to a $\SU(2)$ rotation as explained above, so that the initial spinors $w^s$ and $w^t$ are in fact related by a $\SL(2,\C)$ transformation.

It is possible to directly write this $\SL(2,\C)$ group element mapping $w^s$ to $w^t$. First, we act with the $\SU(2)$ group element $g_{w}$ rotating $w^s$ to $w^t$ up to a norm factor:
\be
g_{w}
=
\f{|w^t\ra [ w^s|-|w^t]\la w^s|}{\sqrt{\la w^s|w^s\ra\la w^t|w^t\ra}}
\,\, \in
 \SU(2)\,,
\quad
g_{w}|w^s]
=
-\lambda^{-1}
\,
|w^t\ra\,,
\quad
g_{w}|w^s\ra
=
+\lambda^{-1}
\,
|w^t]
\,,\quad\textrm{with}\quad
\lambda
=\sqrt{\f{\la w^t|w^t \ra}{\la w^s|w^s \ra}}\,\in\R
\,.
\ee
%
%\be
%\Lambda=\f{|w^t\ra[ w^t|-|w^s]\la w^s|}{\la w^s|w^t\ra} \,\in\SL(2,\C)
%\,,
%\qquad
%\left|\begin{array}{rcl}
%\Lambda|w^s] 
%&=&
%|w^t\ra
%\vspace*{1mm}\\
%(\Lambda^\dagger)^{-1}|w^s\ra
%&=&
%-|w^t]
%\end{array}\right.
%\ee
%
Then we combine this with the $\SL(2,\C)$ dilatation by the factor $\lambda$ in the orthonormal basis $(|w^t\ra,|w^t])$:
\be
D=\f{\lambda |w^t\ra\la w^t|+\lambda^{-1}|w^t][w^t|}{\la w^t|w^t \ra}
\,,\quad
D\, |w^t\ra=\lambda |w^t\ra
\,,\quad
D\, |w^t]=\lambda^{-1} |w^t]\,.
\label{normfactor1}
\ee
Since the norm factor $\lambda$ is real, $D=D^\dagger$ is Hermitian and defines a pure boost in $\SL(2,\C)$. Putting the $\SU(2)$ transformation together with this pure boost gives us the Cartan decomposition of the $\SL(2,\C)$ transformation $\Lambda=Dg_{w}$ mapping $w^s$ to $w^t$:
\be
\Lambda=Dg_{w}
=
\f{|w^t\ra [ w^s|}{\la w^s|w^s\ra}
-
\f{|w^t]\la w^s|}{\la w^t|w^t\ra}
\in\SL(2,\C)
\,,\qquad
\Lambda\,|w^s]=|w^t\ra
\,,\quad
(\Lambda^\dagger)^{-1}\,|w^s\ra=|w^t]
\,.
\ee
Finally, this gives us the whole map from the source spinor $z^s$ to the target spinor $z^t$  as a $\SL(2,\C)$ group element instead of a $\SU(2)$ holonomy as in the original twisted geometry:
\be
G=g^t\Lambda g^s
\,,\qquad
\left|\begin{array}{rcl}
G|z^s\ra 
&=&
|z^t]
\vspace*{1mm}\\
(G^\dagger)^{-1}|z^s]
&=&
-|z^t\ra
\end{array}\right.
\ee
This $\SL(2,\C)$ group element Poisson-commutes with the intermediate closure constraint $\cC$ given in \eqref{closure} and, equivalently, is invariant under $\SU(2)$ gauge transformation at the intermediate vertex linking the loop to the edge.

The trace $\tr\,GG^\dagger=\tr\,\Lambda\Lambda^\dagger=\tr\,DD^\dagger$ gives a measure of the closure defect along the edge and how far the $\SL(2,\C)$ group element $G$ is from the $\SU(2)$ subgroup. Indeed, $\tr\,GG^\dagger$ is always larger or equal to 2 and is equal to 2 if and only if $G$ is a $\SU(2)$ group element\footnotemark.
\footnotetext{We use the Iwasawa decomposition to write $G\in\SL(2,\C)$ generically as the product of a upper complex matrix and a $\SU(2)$ group element:
\be
G= \mat{cc}{\mu & \zeta \\ 0 & \mu^{-1}}g
\quad\textrm{with}\quad
\mu>0
\,,\,\,
\zeta\in\C
\,,\,\,
g\in\SU(2)
\,,\qquad
\tr\,GG^\dagger=\mu^2+\mu^{-2}+|\zeta|^2\,,
\nn
\ee
so that we always have $\tr\,GG^\dagger\ge 2$ and that the equality $\tr\,GG^\dagger= 2$ holds if and only if $\mu=1$ and $\zeta=0$, i.e. if $G=g\in\SU(2)$.
}
We can compute:
\be
\tr\,GG^\dagger-2
=
\tr\,DD^\dagger-2
=
\lambda^2+\lambda^{-2}-2
=
\big{(}
\lambda-\lambda^{-1}
\big{)}^2
\,\,
\ge 0
\,.
\label{normfactor2}
\ee
Thus the trace $\tr\,GG^\dagger$ of the (squared)  $\SL(2,\C)$ group element along the edge gives directly the norm ratio  $\lambda$ between the source and target spinors (and thereby flux vectors) on the coarse-grained edge. If the holonomy $h$ carried by the loop is trivial, $h=\id$, then the norm factor $\lambda=1$  and the trace $\tr\,GG^\dagger=2$, so that the $\SL(2,\C)$ group element $G$ reduces to a $\SU(2)$ holonomy along the edge.

\medskip

To summarize, we consider  the insertion of a loop along a link in a twisted geometry. This loop creates a closure defect from the pont of view the link and relaxes the norm matching constraint between the source and target spinors of the link. Considering the link from a coarse-grained perspective, forgetting about the details of the loop and focusing on the relation between the source and target spinors, this implies that we need to replace the usual $\SU(2)$ holonomy along the link by a $\SL(2,\C)$ group element $G$: the coarse-grained twisted geometry link acquires a boost.
The squared  trace $\tr\,GG^\dagger$ of the $\SL(2,\C)$ element is invariant under $\SU(2)$ gauge transformations and  provides a direct measure of the norm ratio $\la z^s|z^s \ra/\la z^t|z^t \ra$ between the source and target spinors. Other $\SU(2)$-invariant observables attached to the loop, such the Wilson loop $\tr\,h$ or the matrix element $\la\alpha|\beta]=\la\alpha|h|\alpha\ra$ or the scalar product $\vX(w^s)\cdot\vX(w^t)$, are fine data to be forgotten during the coarse-graining process.

This describes the dressing of twisted geometry edges with curvature defects and how twisted geometries become boosted, i.e. acquire $\SL(2,C)$ holonomies instead of the $\SU(2)$ holonomies of the original definition,  during the coarse-graining.

%%%
%\subsection{Compute Poisson brackets !!}
%%%

%%%%%%%%
\section{Dressed Spin Network Edges}
%%%%%%%%

At the quantum level, twisted geometries are quantized into the spin network states of geometry of Loop Quantum Gravity \cite{Dupuis:2011fz,Livine:2011gp,Livine:2013wmq}.
The flux vector norms $|\vX_{e}^s|=|\vX_{e}^t|$ on an edge, or squared-norm of the spinors $\la z_{e}^s|z_{e}^s\ra=\la z_{e}^t|z_{e}^t\ra$, are quantized and  become the spins $j_{e}$ (labelling $\SU(2)$ irreducible representations) carried by the spin network edges.
Here we will see that the insertion of a (little) loop in the middle of an edge creates a curvature defect allowing for different spins at the source and target of the edge, $j_{e}^s=j_{e}^t$. Thus coarse-graining spin networks lead to dressed edges allowing for  a non-trivial propagation of spins. We will further show that spin transitions are mathematically equivalent to $\SL(2,\C)$ boosts along the dressed link, as in the classical case presented in the previous section.

%%%
\subsection{Little Loops as Tadpoles on Edges}
%%%

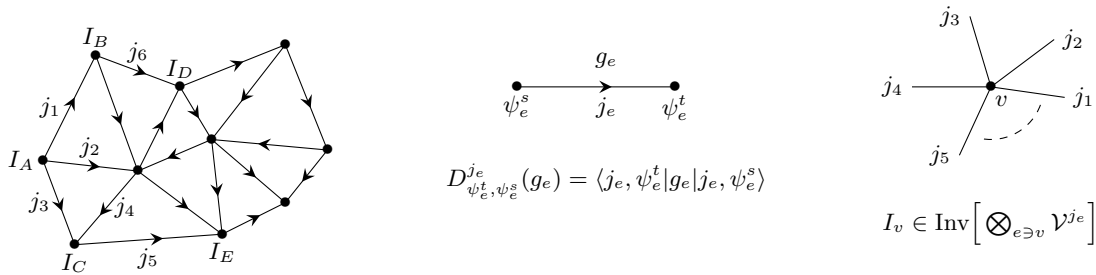
\begin{figure}[h!]

\begin{tikzpicture}[scale=1.4]

\coordinate(a) at (0,0) ;
\coordinate(b) at (.5,1);
\coordinate(c) at (.9,-.1);
\coordinate(d) at (.3,-.8);
\coordinate(e) at (1.3,.7);
\coordinate(f) at (2.3,1.1);
\coordinate(g) at (1.6,.2);
\coordinate(h) at (2.7,.1);
\coordinate(i) at (2.3,-.4);
\coordinate(j) at (1.7,.-.7);

\draw (a) node {$\bullet$} node[left]{$I_{A}$};
\draw (b) node {$\bullet$}node[above]{$I_{B}$};
\draw (c) node {$\bullet$} ;
\draw (d) node {$\bullet$} node[below]{$I_{C}$};
\draw (e) node {$\bullet$} node[above]{$I_{D}$};
\draw (f) node {$\bullet$} ;
\draw (g) node {$\bullet$};
\draw (h) node {$\bullet$};
\draw (i) node {$\bullet$};
\draw (j) node {$\bullet$}node[below]{$I_{E}$};

\alink{a}{b}{left}{j_{1}};
\alink{a}{c}{above}{j_{2}};
\alink{a}{d}{left}{j_{3}};
\link{b}{c};
\alink{c}{d}{right}{j_{4}};
\alink{b}{e}{above}{j_{6}};
\link{e}{f};
\link{c}{e};
\link{e}{g};
\link{f}{g};
\link{f}{h};
\link{h}{g};
\link{g}{c};
\link{h}{i};
\link{g}{i};
\link{j}{i};
\alink{d}{j}{below}{j_{5}};
\link{c}{j};
\link{g}{j};

\coordinate(A) at (4.5,0.7);
\coordinate(B) at (6,0.7);
\link{A}{B};
\draw(5.35,0.5) node{$j_{e}$};
\draw(5.35,0.95) node{$g_{e}$};
\draw (A) node{$\bullet$} node[below]{$\psi^s_{e}$};
\draw (B) node{$\bullet$} node[below]{$\psi^t_{e}$};
\draw(5.35,-0.2) node{$D^{j_{e}}_{\psi_{e}^t,\psi_{e}^s}(g_{e})=\la j_{e}, \psi_{e}^t|g_{e}|j_{e}, \psi_{e}^s\ra$};

\coordinate(O) at (9,0.7);
\draw (O) node{$\bullet$} ++(0.1,0) node[below]{$v$};
\draw (O) --++(0.7,-0.1) node[right]{$j_{1}$};
\draw (O) --++(0.6,0.45) node[right]{$j_{2}$};
\draw (O) --++(-0.2,0.67) node[left]{$j_{3}$};
\draw (O) --++(-0.75,0) node[left]{$j_{4}$};
\draw (O) --++(-0.3,-0.65) node[left]{$j_{5}$};
\centerarc[dashed](O)(-20:-100:0.5);
\draw(9,-0.6) node{$I_{v}\in\textrm{Inv}\Big{[}\bigotimes_{e\ni v}\cV^{j_{e}}\Big{]}$};

\end{tikzpicture}

\caption{On the left, a spin network, on a closed oriented graph $\Gamma$, is a basis state of 3d quantum geometry. It is labeled with spins $j_{e}\in\f N2$ on the graph edges and intertwiner states  on the graph vertices. In the middle, the edges carry $\SU(2)$ group elements $g_{e}$, or holonomies, which are considered as matrices in the representation given the edge spin $j_{e}$. Finally, on the right,  intertwiners at the  vertices are $\SU(2)$-invariant tensors -or equivalently singlet states- in the tensor product of the spins living on the edge around the vertex. These intertwiners allow to glue together the matrix elements of the $\SU(2)$ holonomies into a $\SU(2)$-invariant scalar defining the spin network wave-function.}
\label{fig:spinnetwork}
\end{figure}

Let us consider a spin network state living on a graph $\Gamma$ and focus on an edge $e\in\Gamma$. In a pure state, the edge carries a spin $j_{e}\in\f\N2$ determining an irreducible unitary representation of the $\SU(2)$ Lie group, as well as two states in that representation attached to the source and target of the edge, $|j_{e}, \psi_{e}^s\ra$ and $|j_{e}, \psi_{e}^t\ra$ both in the Hilbert space $V^{j_{e}}$ carrying the $\SU(2)$-representation of spin $j_{e}$.
%$|j_{e}, m_{e}^s\ra$ and $|j_{e}, m_{e}^t\ra$
%
This data defines a function of a $\SU(2)$ group element $g_{e}$, given by the corresponding Wigner matrix element $D^{j_{e}}_{\psi_{e}^t,\psi_{e}^s}(g_{e})=\la j_{e}, \psi_{e}^t|g_{e}|j_{e}, \psi_{e}^s\ra$. This is the part of the spin network wave-function corresponding to the edge $e$. As illustrated on figure \ref{fig:spinnetwork}, these matrix elements are combined with intertwiners (or $\SU(2)$-invariant tensors) living at the graph vertices $v\in\Gamma$ to form the overall spin network wave-function as a function of $\SU(2)$ group elements $g_{e}\in\SU(2)$ living on every edge of the graph $e\in\Gamma$.
The interested reader can find short reviews of spin networks and their algebraic structure in e.g. \cite{Livine:2013gna,Bodendorfer:2016uat}.

\begin{figure}[h!]

\begin{subfigure}[t]{.5\linewidth}
\begin{tikzpicture}[scale=1.2]

\coordinate(a) at (0,0);
\coordinate(b) at (1,0);
\coordinate(c) at (2,0);

\draw (a) node {$\bullet$};
\draw (c) node {$\bullet$};
%\draw (b) node {{\large $\bullet$}};
%\draw[fill=lightgray] (b) circle(0.1);
%\draw[fill=black] (b) circle(0.07);

\alink{a}{b}{below}{j^s};
\alink{b}{c}{below}{j^t};

\draw[scale=4,decoration={markings,mark=at position 0.52 with {\arrow[scale=1.5,>=stealth]{>}}},postaction={decorate}] (b) to[in=+45,out=+135,loop]  node[midway,above]{$k$}(b) ;

\fill[gray] (b) circle(0.07);

\coordinate(A) at (4,-0.2);
\coordinate(B) at (5,-0.2);
\coordinate(C) at (6,-0.2);
\coordinate(D) at (5,.5);

\draw (A) node {$\bullet$};
%\draw (B) node {{\tiny $\bullet$}};
\draw (C) node {$\bullet$};
%\draw (D) node {{\tiny $\bullet$}};

\alink{A}{B}{below}{j^s};
\alink{B}{C}{below}{j^t};
\draw[dotted,thick] (B)--(D) node[midway,right]{$J$};

\draw[scale=2.5,decoration={markings,mark=at position 0.53 with {\arrow[scale=1.3,>=stealth]{>}}},postaction={decorate}] (D) to[in=+45,out=+135,loop]  node[midway,above]{$k$}(D) ;

\fill[gray] (B) circle(0.05);
\fill[gray] (D) circle(0.05);

\draw[decoration={markings,mark=at position 1 with {\arrow[scale=1.5,>=latex]{>}}},postaction={decorate}] (2.3,.6)--(3.7,.6) ;

\end{tikzpicture}

\caption{The 4-valent intertwiner can be unfolded into two 3-valent intertwiners linked by a virtual edge carrying an arbitrary spin $J$. The  3-valent vertices contribute the corresponding Clebsch-Gordan coefficients, while the intermediate spin $J$ is necessarily an integer to respect the parity of the spin recoupling with the spin $k$ carried by the loop.}

\end{subfigure}
\hspace*{10mm}
\begin{subfigure}[t]{.4\linewidth}
\centering
\begin{tikzpicture}
\coordinate(A) at (0,-0.2);
\coordinate(B) at (1,-0.2);
\coordinate(C) at (2.5,-0.2);
\coordinate(D) at (1,.5);

\draw (A) node {$\bullet$};
%\draw (B) node {{\tiny $\bullet$}};
\draw (C) node {$\bullet$};
%\draw (D) node {{\tiny $\bullet$}};

\link{A}{B};
\draw(.5,-0.5) node{$j^s$};
%\alink{B}{C}{below}{j^t=j^s+a};
\link{B}{C};
\draw(2.5,-0.5) node{$j^t=j^s+a$};
\draw[dotted,thick] (B)--(D) node[midway,right]{$J=1$};

\draw[scale=2.8,decoration={markings,mark=at position 0.53 with {\arrow[scale=1.3,>=stealth]{>}}},postaction={decorate}] (D) to[in=+45,out=+135,loop]  node[midway,above]{$k=\f12$}(D) ;

\fill[gray] (B) circle(0.06);
\fill[gray] (D) circle(0.06);

\draw (3.5,0.6) node{with $a=0,\pm1$};

\end{tikzpicture}

\caption{The simplest case of a loop attached to a spin network edge is a loop spin $k=\f12$ linked to the edge by an intermediate spin $J=1$. This defines a fundamental curvature excitation along the edge. It produces a spin shift along the edge $(j^t-j^s)=a\in\{0,\pm1\}$.}

\end{subfigure}

\caption{The decomposition in the spin basis of a 4-valent intertwiner attaching the loop to the edge}
\label{fig:tadpole}
\end{figure}
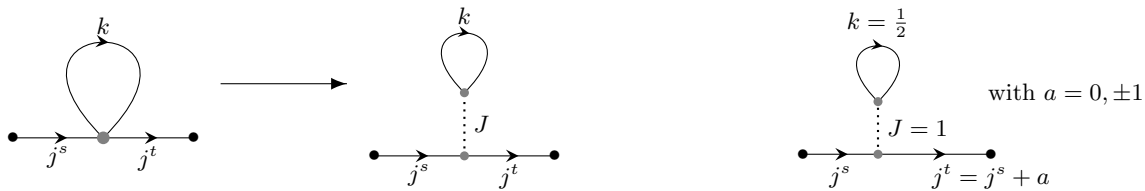

Here we would like to describe the effect of inserting a loop along an edge $e$. As drawn on figure \ref{fig:tadpole}, the insertion of the loop creates a 4-valent vertex linking the loop to the original edge. A basis of spin network states for such a configuration is labeled by a spin $k\in\f\N2$ carried by the loop, an intertwiner basis state at the intermediate 4-valent vertex which we can unfold into  two 3-valent vertices linked by a spin $J\in\N$ channel and spins $j_{e}^s,j_{e}^t$ living on the edge $e$.
The key effect is that the intermediate vertex allows the edge to carry different spins, $j_{e}^s$ and $j_{e}^t$, at its source and target vertices. This implements the quantum counterpart of the non-trivial norm factor in classical twisted geometries, as in equations \eqref{normfactor1} and \eqref{normfactor2}.

Diagrammatically, the spin network edge with the loop insertion looks very much like a Feynman diagram propagator renormalized by a tadpole contribution. Its effect is similar, it creates a non-trivial propagation of the spin, thereby of quanta of area, along the edge. This non-trivial area propagator reflects the non-trivial $\SU(2)$ holonomy, i.e. curvature, that can develop around the loop.

%%%
\subsection{Clebsch-Gordan Coefficients as Spin Propagator}
%%%

Let us study in more details the simplest loop insertion, with a loop carrying a spin $k=\f12$  corresponding to the fundamental $\SU(2)$-representation. By the triangular inequalities of spin recoupling, it can linked by the edge by a spin $J=0$ or a spin $J=1$. The trivial link with $J=0$ would correspond to a decoupled loop having no effect on the edge. We thus assume that the $\f12$-spin loop is linked to the edge by a spin $J=1$, as illustrated on figure \ref{fig:tadpole}. This can be considered as defining the fundamental excitation of curvature along a spin network edge.

The loop insertion changes the spin network wave-function, more precisely its part corresponding to the considered edge, by the insertion of a Clebsch-Gordan coefficient sandwiched in between the  $\SU(2)$ holonomies living on each half-edge. Dropping the edge index $e$, this reads:
\be
\la j, \psi^t|g^tg^s|j, \psi^s\ra
\quad
\longrightarrow
%\longmapsto
\quad
\la j^t, \psi^t|g^t|j^t,m^t\ra\,
C^{j^s J\,|j^t}_{m^s M\,|m^t}
\,\la j^s,m^s|g^s|j^s, \psi^s\ra\,\Theta_{J,M}^{(k)}(h)
\,,
\ee
where the Clebsch-Gordan coefficients recouples the edge spins $j^{s,t}$ with the tadpole spin $J$ attaching the loop to the edge:
\be
C^{j^s J\,|j^t}_{m^s M\,|m^t}
\equiv
\la j^t ,m^t\,|\,(j^s,m^s),(J,M) \ra
\,,
\ee
vanishing unless $m^t=m^s+M$, and the function $\Theta_{J,M}^{(k)}(h)$ accounting for the modulating effect of the $\SU(2)$ holonomy living around the loop:
\be
\Theta_{J,M}^{(k)}(h)
\,=\,
\la (J,M),(k,m)\,|\,k,\tm\ra
\,
\la k,\tm |h| k,m\ra\,.
\ee
Here we have used the standard basis $|j,m\ra$ for the Hilbert spaces $V^j$ carrying the spin-$j$ representation labeled by the magnetic moment $m$ running in integer from $-j$ to $+j$.

In the special case with $k=\f12$ and $J=1$, we can compute exactly the Clebsh-Gordan coefficients entering the $\Theta$-factors. With $M$ running from $-1$ to $+1$ and the spin-$\f12$ basis vectors written as $|\up\ra$ and $|\down\ra$, the only non-vanishing symbols are (dropping the indices $J$ and $k$):
\be
\begin{array}{lcl}
\la 1,\down|\up\ra &=& +\sqrt{\f23} \\
\la -1,\up|\down\ra &=& -\sqrt{\f23} \\
\end{array}
\qquad
\begin{array}{lcl}
\la 0,\up|\up\ra &=& -\sqrt{\f13} \\
\la 0,\down|\down\ra &=& +\sqrt{\f13} \\
\end{array}
\nn
\ee
so that the $\Theta$-factors are given by the matrix elements of the $\SU(2)$ group element $h$ living on the loop:
\be
\Theta_{1}(h)
=
\sqrt{\f23}\,\la \up|h|\down\ra
\,,\quad
\Theta_{-1}(h)
=
-\sqrt{\f23}\,\la \down|h|\up\ra
\,,\quad
\Theta_{0}(h)
=
\sqrt{\f13}\,\big{[}
\la \down|h|\down\ra-\la \up|h|\up\ra
\big{]}
\,,
\ee
which are respectively the trace of the contraction of $h$ with the Pauli matrices $\sigma_{-}$, $\sigma_{+}$ and $\sigma_{3}$.
For instance, if $h=e^{i\theta\sigma_{3}}$, then only the factor $\Theta_{0}$ does not vanish and is equal to $\Theta_{0}(e^{i\theta\sigma_{3}})=-2\sin\theta/\sqrt3$. Then if we rotate $e^{i\theta\sigma_{3}}$ by acting on it by conjugation to reach an arbitrary $\SU(2)$ group element $h$, the rotation acts on $M$ (as a 3d rotation in the spin-$1$ representation).

This shows that the effect of the $\Theta$-factors  is to modulate the contribution of each moment $M$ to the spin network wave-function depending on the value of the loop holonomy $h$. The more interesting effect, to our point of view, is due to the insertion of the Clebsh-Gordan coefficients $C^{j^s J\,|j^t}_{m^s M\,|m^t}$, which allows for spin shifts along the edge and leading to a target spin $j^t$ possibly different from the source spin $j^s$.

Below we show that those coefficients in the $J=1$ case, $C^{j^s, 1\,|j^t}_{m^s M\,|m^t}$ with $M=-1,0,+1$ can be generated by the action of $\SL(2,\C)$ boost generators, thereby upgrading the $\SU(2)$ holonomy on the edge with boost insertions.

%%%
\subsection{Spin Transitions from Lorentz boosts}
\label{CBsl2c}
%%%

Let us then focus on the spin transition rendered possible by the tadpole insertion. Tensoring the spin $j^s$ at the edge source with the  spin $J$ on the link with the loop leads to a target spin $j^t$ possibly ranging from $|j^s-J|$ to $j^s+J$ according to the triangular inequalities of spin recoupling.
In the simplest case, with $J$ set to 1, the target spin $j^t=j^s+a$ can differ from the source spin $j^s$ by at most 1, therefore with $a\in\{0,\pm1\}$. We can organize the relevant Clebsch-Gordan coefficients $\la j+a,m+M \,|\, (j,m),(1,M)\ra$, with $(j^s,m^s)=(j,m)$ and $(j^t,m^t)=(j+a,m+M)$, in the table below:
\be
%\begin{tabular}{|c||c|c|c|}
\la j+a,m+M \,|\, (j,m),(1,M)\ra
\quad=\quad
\begin{tabular}{|c||*{3}{c|}}
\hline
\backslashbox{a}{M}
&\makebox[3em]{-1}&\makebox[3em]{0}&\makebox[3em]{+1}
\\\hline\hline
\rule{0pt}{16pt}
-1
& $\sqrt{\f{(j+m)(j+m-1)}{2j(2j+1)}}$
& $-\sqrt{\f{(j-m)(j+m)}{j(2j+1)}}$
& $\sqrt{\f{(j-m)(j-m-1)}{2j(2j+1)}}$
\\
\hline
\rule{0pt}{16pt}
0
& $\sqrt{\f{(j+m)(j-m+1)}{2j(j+1)}}$
& \raisebox{4pt}[0pt][0pt]{$ \f{m}{\sqrt{j(j+1)}}$}
& $-\sqrt{\f{(j-m)(j+m+1)}{2j(j+1)}}$
\\
\hline
\rule{0pt}{16pt}
+1
& $\sqrt{\f{(j-m+1)(j-m+2)}{2(j+1)(2j+1)}}$
& $\sqrt{\f{(j-m+1)(j+m+1)}{(j+1)(2j+1)}}$
& $\sqrt{\f{(j+m+1)(j+m+2)}{2(j+1)(2j+1)}}$
\\
\hline
\end{tabular}
\ee
It turns out that these Clebsch-Gordan coefficients can be derived from the action of $\SL(2,\C)$ boost generators $\vK$.
Let us consider the unitary representations of the $SL(2,\C)$ Lie group. They are labeled by two numbers $(n,\rho)$, with the principal series of irreducible unitary representation given by $n\in\f{\N}2$ and $\rho\in\R$ and the supplementary series defined by $n=0$ and $\rho\in i\R$, $|\rho|<1$. These representations can be decomposed in representations of the $\SU(2)$ subgroup in $\SL(2,\C)$ (which is its maximal compact subgroup). Calling $\cR^{(n,\rho)}$ the Hilbert space carrying the $(n,\rho)$ unitary $\SL(2,\C)$-representation, it is the direct sum of all $\SU(2)$-representations with spins $j$ greater than $n$ by integer steps:
\be
\cR^{(n,\rho)}=\bigoplus_{j\in n+\N}\cV^j
\,.
\ee
Natural basis states for the Hilbert space $\cR^{(n,\rho)}$ are thus $|(n,\rho)\,j,m\ra$, labeled by the $\sl(2,\C)$ labels $(n,\rho)$ and by the $\su(2)$ basis labels $(j,m)$.
We distinguish the $\su(2)$ generators $J_{i}$ from the boost generators $K_{i}$ in the $\sl(2,\C)$ Lie algebra, defined with the commutation relations:
\beq
&[J_{3},J_{\pm}]=\pm J_{\pm}
\,,\quad
[J_{+},J_{-}]=2J_{3}
\,,
&
\\
&
[J_{3},K_{\pm}]=\pm  K_{\pm}
\,,\quad
[J_{3},K_{3}]=0
\,,\quad
&
\nn\\
&
[K_{3},J_{\pm}]=\pm  K_{\pm}
\,,\quad
[K_{3},K_{\pm}]=\mp  J_{\pm}
\,,\quad
&
\nn\\
&
[J_{+},K_{-}]=[K_{+},J_{-}]=2  K_{3}
\,,\quad
[K_{+},K_{-}]=-2J_{3}
\,,\quad
[J_{\pm},K_{\pm}]=0
\,.
&\nn
\eeq
The $\su(2)$ generators $J_{i}$ act in each $\cV^j$ space, while the boost generators $K_{i}$ creates transitions between spins $j$. Their detailed action  is given in appendix \ref{sl2C}.

Here we are interested in the relation between the action of the boost generator and the Clebsch-Gordan coefficients with $J=1$, summarized in the equalities:
 \be
 \label{boostshift}
 \la (n,\rho)\,j+a,m+M|K_{M}|(n,\rho)\,j,m\ra
 =
 \,
v_{M}\, \gamma_{a}^{(n,\rho)}[j]\,
 \la j+a,m+M|(j,m)(1,M)\ra
 \,,
 \quad\textrm{with}\quad
\left|\begin{array}{ll}
v_{0}&=1 \\
v_{-}&=+\sqrt{2}\\
v_{+}&=-\sqrt{2}
\end{array}\right.
\,,
 \ee
 where the $v_{M}$'s are normalization factors for the boost generators and the factors $\gamma_{a}^{(n,\rho)}[j]$ depend on the spin shift $a$, the $\SL(2,\C)$ representation $(n,\rho)$ and the spin $j$, but crucially do not depend on the state label $m$:
 \be
 \gamma_{0}^{(n,\rho)}[j]
 =
 \f{n\rho}{\sqrt{j(j+1)}}
 \,,\qquad
 \gamma_{-}^{(n,\rho)}[j]
 =
 -i\,\sqrt{\f{(j^2-n^2)(j^2+\rho^2)}{j(2j-1)}}
 \,,\qquad
\gamma_{+}^{(n,\rho)}[j]
=
\gamma_{-}^{(n,\rho)}[j+1]
\,.
 \ee
 The factor $(j^2-n^2)$ ensures that the spin $j$ always remains larger or equal to the $\sl(2,\C)$ representation integer label $n$. These factors control how the choice of the $\sl(2,\C)$ representation $(n,\rho)$ modulates the spin shift $j^s=j\rightarrow\,j^t=j+a$.
This key equality \eqref{boostshift} works because the boost generators $K_{i}$ are vector operators under the action of the $\SU(2)$ Lie group.
 
 An interesting special case is given by so-called ``simple representations'' $(0,\rho)$ appearing for instance in the Barrett-Crane state-sum models for 3+1-dimension quantum gravity (formulated as a quasi-topological field theory) \cite{Barrett:1997gw,Barrett:1999qw}. In that case, boosts necessary shift the spins since the coefficient $\gamma_{0}$ vanishes (for $a=0$) and they can shift the spins all the way down to $j=0$ or up towards $\infty$:
  \be
 \gamma_{0}^{(0,\rho)}[j]
 =
0
 \,,\qquad
\gamma_{-}^{(0,\rho)}[j]
=
-i\,\sqrt{\f{j(j^2+\rho^2)}{(2j-1)}}
\,.
 \ee
 
 \medskip
 
 This shows that we can switch a fundamental curvature excitation along the edge defined by a loop attached with the smallest possible spin -the spin 1- with the insertion of a boost generator $K_{i}$ along the edge. They create similar spin shifts between the source and target spins of the spin network edge. The  distribution modulating the spin shift as defined by the loop (the spin $k$  and the $\SU(2)$ holonomy $h$ carried by the loop) can be mapped onto a distribution over the choice of $\sl(2,\C)$ representations\footnotemark{} $(n,\rho)$.
\footnotetext{
We focused over unitary $\SL(2,\C)$-representations in order to facilitate the comparaison with the existing spinfoam models for 3+1-dimensional quantum gravity, but this restriction to unitary representations is not a mathematical requirement and equation \eqref{boostshift} also holds for non-unitary representations when $n$ and $\rho$ are both arbitrary complex numbers.
}

This embedding of $\SU(2)$ spins into $\SL(2,\C)$ unitary representations, or equivalently the dressing of a spin network with $\SL(2,\C)$ representations, is reminiscent of the projective spin network structures \cite{Livine:2002ak,Dupuis:2010jn,Ding:2010ye}, used as basis states for spinfoam path integral models for quantum gravity in 3+1 spacetime dimensions of the Barrett-Crane type \cite{Barrett:1997gw,Barrett:1999qw} or EPRL-FK type \cite{Engle:2007wy,Freidel:2007py,Livine:2007ya,Rovelli:2010vv,}. Although projective spin networks were introduced for a different reason than coarse-graining curvature excitations, in order to account for the embedding of the 3d spin network geometry into the surrounding space-time, the mathematical effect of $\SL(2,\C)$ boosts creating spin shifts along spin network edges is exactly the same. Nevertheless, this possibility has usually not been investigated in the spinfoam framework and typical models systematically project on the $j^s=j^t$ sector for every spin network edge \cite{Dupuis:2010jn,Ding:2010ye}. The work presented here provides a physical interpretation of the sector $j^s\ne j^t$ and motivation to explore for it -better understand the coarse-graining of spin network and how to renormalize their structure to account for microscopic curvature excitations.

%%%
\subsection{Spin propagation and spin waves}
\label{spinwave}
%%%

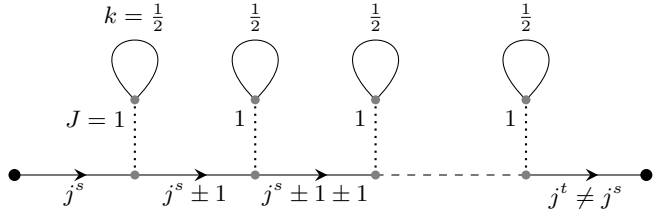
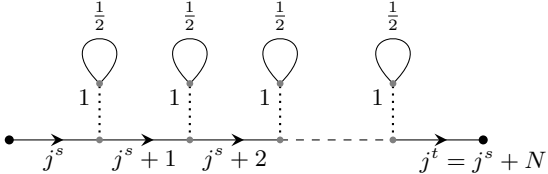
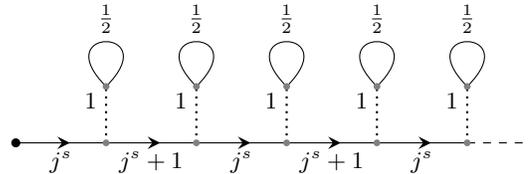
\begin{figure}[h!]

\begin{subfigure}[t]{\linewidth}
\centering
\begin{tikzpicture}[scale=2]

\coordinate(S) at (0,0);
\coordinate(T) at (4.2,0);

\coordinate(v1) at (0.8,0);
\coordinate(v2) at (1.6,0);
\coordinate(v3) at (2.4,0);
\coordinate(vN) at (3.4,0);

%\link{S}{v1};
%\link{v1}{v2};
%\link{v2}{v3};
%\link{vN}{T};
\alink{S}{v1}{below}{j^s};
\alink{v1}{v2}{below}{j^s\pm 1};
\alink{v2}{v3}{below}{j^s\pm 1\pm 1};
\alink{vN}{T}{below}{j^t\ne j^s};
\draw[dashed] (v3)--(vN);

%\draw(.5,-0.5) node{$j^s$};

\path
let
\p1 =($(v1)+(0,0.5)$),
\p2 =($(v2)+(0,0.5)$),
\p3 =($(v3)+(0,0.5)$),
\p4 =($(vN)+(0,0.5)$)
in
coordinate (V1) at (\p1)
coordinate (V2) at (\p2)
coordinate (V3) at (\p3)
coordinate (VN) at (\p4);

\draw[dotted,thick](v1)--(V1) node[near end,left]{$J=1$};
\draw[dotted,thick](v2)--(V2) node[near end,left]{1};
\draw[dotted,thick](v3)--(V3) node[near end,left]{1};
\draw[dotted,thick](vN)--(VN) node[near end,left]{1};

\draw[scale=1.5] (V1) to[in=+45,out=+135,loop]  node[midway,above]{$k=\f12$}(V1) ;
\draw[scale=1.5] (V2) to[in=+45,out=+135,loop]  node[midway,above]{$\f12$}(V2) ;
\draw[scale=1.5] (V3) to[in=+45,out=+135,loop]  node[midway,above]{$\f12$}(V3) ;
\draw[scale=1.5] (VN) to[in=+45,out=+135,loop]  node[midway,above]{$\f12$}(VN) ;

\fill[black] (S) circle(0.04);
\fill[black] (T) circle(0.04);

\fill[gray] (v1) circle(0.03);
\fill[gray] (v2) circle(0.03);
\fill[gray] (v3) circle(0.03);
\fill[gray] (vN) circle(0.03);

\fill[gray] (V1) circle(0.03);
\fill[gray] (V2) circle(0.03);
\fill[gray] (V3) circle(0.03);
\fill[gray] (VN) circle(0.03);

\end{tikzpicture}

\caption{Spin propagation along a spin network edge: each loop insertion $k=\f12$ taken in the tadpole basis with an linking spin $J=1$ creates a possible spin shift of at most $\pm 1$, leading to a non-trivial propagation of the spin along the edges with possibly the target spin $j^t$ different from the source spin $j^s$}

\end{subfigure}
%\hspace*{10mm}
%
\begin{subfigure}[t]{.4\linewidth}
\begin{tikzpicture}[scale=1.5]

\coordinate(S) at (0,0);
\coordinate(T) at (4.2,0);

\coordinate(v1) at (0.8,0);
\coordinate(v2) at (1.6,0);
\coordinate(v3) at (2.4,0);
\coordinate(vN) at (3.4,0);

%\link{S}{v1};
%\link{v1}{v2};
%\link{v2}{v3};
%\link{vN}{T};
\alink{S}{v1}{below}{j^s};
\alink{v1}{v2}{below}{j^s+ 1};
\alink{v2}{v3}{below}{j^s+2};
\link{vN}{T};
\draw[dashed] (v3)--(vN);
\draw (T) node[below]{$j^t=j^s+N$};

%\draw(.5,-0.5) node{$j^s$};

\path
let
\p1 =($(v1)+(0,0.5)$),
\p2 =($(v2)+(0,0.5)$),
\p3 =($(v3)+(0,0.5)$),
\p4 =($(vN)+(0,0.5)$)
in
coordinate (V1) at (\p1)
coordinate (V2) at (\p2)
coordinate (V3) at (\p3)
coordinate (VN) at (\p4);

\draw[dotted,thick](v1)--(V1) node[near end,left]{1};
\draw[dotted,thick](v2)--(V2) node[near end,left]{1};
\draw[dotted,thick](v3)--(V3) node[near end,left]{1};
\draw[dotted,thick](vN)--(VN) node[near end,left]{1};

\draw[scale=1.5] (V1) to[in=+45,out=+135,loop]  node[midway,above]{$\f12$}(V1) ;
\draw[scale=1.5] (V2) to[in=+45,out=+135,loop]  node[midway,above]{$\f12$}(V2) ;
\draw[scale=1.5] (V3) to[in=+45,out=+135,loop]  node[midway,above]{$\f12$}(V3) ;
\draw[scale=1.5] (VN) to[in=+45,out=+135,loop]  node[midway,above]{$\f12$}(VN) ;

\fill[black] (S) circle(0.04);
\fill[black] (T) circle(0.04);

\fill[gray] (v1) circle(0.03);
\fill[gray] (v2) circle(0.03);
\fill[gray] (v3) circle(0.03);
\fill[gray] (vN) circle(0.03);

\fill[gray] (V1) circle(0.03);
\fill[gray] (V2) circle(0.03);
\fill[gray] (V3) circle(0.03);
\fill[gray] (VN) circle(0.03);

\end{tikzpicture}

\caption{Spin diffusion along a spin network edge as an example of spin propagation: starting from the source spin $j^s$, each loop insertion increments the spin by $+1$.}
\label{fig:spindiffusion}

\end{subfigure}
\hspace*{20mm}
\begin{subfigure}[t]{.4\linewidth}
\centering
\begin{tikzpicture}[scale=1.5]

\coordinate(S) at (0,0);
\coordinate(T) at (4.5,0);

\coordinate(v1) at (0.8,0);
\coordinate(v2) at (1.6,0);
\coordinate(v3) at (2.4,0);
\coordinate(v4) at (3.2,0);
\coordinate(v5) at (4,0);

%\link{S}{v1};
%\link{v1}{v2};
%\link{v2}{v3};
%\link{vN}{T};
\alink{S}{v1}{below}{j^s};
\alink{v1}{v2}{below}{j^s+ 1};
\alink{v2}{v3}{below}{j^s};
\alink{v3}{v4}{below}{j^s+ 1};
\alink{v4}{v5}{below}{j^s};

\draw[dashed] (v5)--(T);

%\draw(.5,-0.5) node{$j^s$};

\path
let
\p1 =($(v1)+(0,0.5)$),
\p2 =($(v2)+(0,0.5)$),
\p3 =($(v3)+(0,0.5)$),
\p4 =($(v4)+(0,0.5)$),
\p5 =($(v5)+(0,0.5)$)
in
coordinate (V1) at (\p1)
coordinate (V2) at (\p2)
coordinate (V3) at (\p3)
coordinate (V4) at (\p4)
coordinate (V5) at (\p5);

\draw[dotted,thick](v1)--(V1) node[near end,left]{1};
\draw[dotted,thick](v2)--(V2) node[near end,left]{1};
\draw[dotted,thick](v3)--(V3) node[near end,left]{1};
\draw[dotted,thick](v4)--(V4) node[near end,left]{1};
\draw[dotted,thick](v5)--(V5) node[near end,left]{1};

\draw[scale=1.5] (V1) to[in=+45,out=+135,loop]  node[midway,above]{$\f12$}(V1) ;
\draw[scale=1.5] (V2) to[in=+45,out=+135,loop]  node[midway,above]{$\f12$}(V2) ;
\draw[scale=1.5] (V3) to[in=+45,out=+135,loop]  node[midway,above]{$\f12$}(V3) ;
\draw[scale=1.5] (V4) to[in=+45,out=+135,loop]  node[midway,above]{$\f12$}(V4) ;
\draw[scale=1.5] (V5) to[in=+45,out=+135,loop]  node[midway,above]{$\f12$}(V5) ;

\fill[black] (S) circle(0.04);
%\fill[black] (T) circle(0.04);

\fill[gray] (v1) circle(0.03);
\fill[gray] (v2) circle(0.03);
\fill[gray] (v3) circle(0.03);
\fill[gray] (v4) circle(0.03);
\fill[gray] (v5) circle(0.03);

\fill[gray] (V1) circle(0.03);
\fill[gray] (V2) circle(0.03);
\fill[gray] (V3) circle(0.03);
\fill[gray] (V4) circle(0.03);
\fill[gray] (V5) circle(0.03);

\end{tikzpicture}

\caption{Spin wave along a spin network edge as an example of spin propagation: the spin oscillates between $j^s$ and $j^s+1$ with each loop insertion.}
\label{fig:spinwave}

\end{subfigure}

\caption{Spin propagation  due to loop insertions representing curvature excitations along a spin network edge.}
\label{fig:spinpropagation}
\end{figure}

Instead of a single loop insertion along an edge, we expect the possible insertion of multiple loops and curvature excitation along every spin network edges. Such sequences of tadpoles, as depicted on figure \ref{fig:spinpropagation}, would lead to a non-trivial propagation of the spin along the edge from its source spin $j^s$ to its target spin $j^t$. Then prescribing a probability amplitude for tadpole insertion would lead to a non-trivial spin propagator along spin network edges, similarly to propagators on a Feynman diagram. We have no proposal yet for such a probability amplitude of curvature excitations. This would definitely be related to a choice of  dynamics for spin networks. At the kinematical level, we can nevertheless describe two types of possible behavior: a diffusive regime and a oscillatory regime.

Let us imagine a sequence of several loop insertions along an edge, all defining fundamental curvature excitations linked to the edge with a spin $J=1$. There are two natural behaviors.
First, the spin shift could accumulate constructively with the spin increasing along the edge, as on the left hand side of fig.\ref{fig:spindiffusion}, $j^s\arr j^s+1 \arr j^s +2 \arr j^s +3\arr\dots$. Such a diffusive configuration can be understood geometrically as the area increasing along the edges, similarly to a conformal transformation of the metric from the source node to the target node.
Second, the spin shift could evolve destructively, with a positive shift followed by a negative shift and so on, thereby creating spin oscillations, as on the right hand side of fig.\ref{fig:spinwave}, $j^s\arr j^s+1 \arr j^s \arr j^s +1\arr\dots$.
Instead of tight oscillations between $j^s$ and $j^s+1$, one can easily imagine waves with a larger amplitude of spin shifts with oscillations spread over several loop insertions.
Since the spin gives the quanta of area, it would be interesting to investigate if such a ``spin wave'' or ``area wave'' can model a gravitational wave or whether it is a different type of quantum gravity excitation. It could also be enlightening for loop quantum gravity if we could classify the various types of such area waves, i.e. classify the types of  probability amplitudes of loop insertions leading to such waves.

%%%
\subsection{Boosting holonomies}
%%%

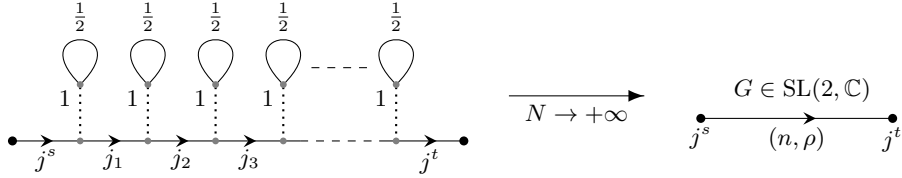
\begin{figure}[h!]

\centering
\begin{tikzpicture}[scale=1.5]

\coordinate(S) at (0,0);
\coordinate(T) at (4,0);

\coordinate(v1) at (0.6,0);
\coordinate(v2) at (1.2,0);
\coordinate(v3) at (1.8,0);
\coordinate(v4) at (2.4,0);
\coordinate(vN) at (3.4,0);

\alink{S}{v1}{below}{j^s};
\alink{v1}{v2}{below}{j_{1}};
\alink{v2}{v3}{below}{j_{2}};
\alink{v3}{v4}{below}{j_{3}};

\draw (v4)--(2.55,0);
\draw[dashed] (2.55,0)--(3.25,0);
\draw[dashed] (2.65,0.65)--(3.15,0.65);
\draw (3.25,0)--(vN);

\alink{vN}{T}{below}{j^t};

%\draw(.5,-0.5) node{$j^s$};

\path
let
\p1 =($(v1)+(0,0.5)$),
\p2 =($(v2)+(0,0.5)$),
\p3 =($(v3)+(0,0.5)$),
\p4 =($(v4)+(0,0.5)$),
\p5 =($(vN)+(0,0.5)$)
in
coordinate (V1) at (\p1)
coordinate (V2) at (\p2)
coordinate (V3) at (\p3)
coordinate (V4) at (\p4)
coordinate (VN) at (\p5);

\draw[dotted,thick](v1)--(V1) node[near end,left]{1};
\draw[dotted,thick](v2)--(V2) node[near end,left]{1};
\draw[dotted,thick](v3)--(V3) node[near end,left]{1};
\draw[dotted,thick](v4)--(V4) node[near end,left]{1};
\draw[dotted,thick](vN)--(VN) node[near end,left]{1};

\draw[scale=1.5] (V1) to[in=+45,out=+135,loop]  node[midway,above]{$\f12$}(V1) ;
\draw[scale=1.5] (V2) to[in=+45,out=+135,loop]  node[midway,above]{$\f12$}(V2) ;
\draw[scale=1.5] (V3) to[in=+45,out=+135,loop]  node[midway,above]{$\f12$}(V3) ;
\draw[scale=1.5] (V4) to[in=+45,out=+135,loop]  node[midway,above]{$\f12$}(V4) ;
\draw[scale=1.5] (VN) to[in=+45,out=+135,loop]  node[midway,above]{$\f12$}(VN) ;

\fill[black] (S) circle(0.04);
\fill[black] (T) circle(0.04);

\fill[gray] (v1) circle(0.03);
\fill[gray] (v2) circle(0.03);
\fill[gray] (v3) circle(0.03);
\fill[gray] (v4) circle(0.03);
\fill[gray] (vN) circle(0.03);

\fill[gray] (V1) circle(0.03);
\fill[gray] (V2) circle(0.03);
\fill[gray] (V3) circle(0.03);
\fill[gray] (V4) circle(0.03);
\fill[gray] (VN) circle(0.03);

\draw[decoration={markings,mark=at position 1 with {\arrow[scale=1.5,>=latex]{>}}},postaction={decorate}] (4.4,0.4)--(5.6,0.4) node[midway,below]{$N\rightarrow +\infty$};

\coordinate (s) at (6.1,0.2);
\coordinate (t) at (7.8,0.2);

\draw (s) node[below]{$j^s$};
\draw (t) node[below]{$j^t$};
\draw (7,0.45) node{$G\in\SL(2,\C)$};

\alink{s}{t}{below}{(n,\rho)};

\fill[black] (s) circle(0.04);
\fill[black] (t) circle(0.04);

\end{tikzpicture}
\caption{A $\SL(2,\C)$ group element $G$ acting along a spin network edge can be decomposed as an infinite sequence of loop insertions along the edge, and vice-versa the continuum limit of a sequence of loop insertions can be interpreted as a superposition of $\SL(2,\C)$ holonomies along the edge. This realizes the equivalence of curvature excitations along a spin network edge with boosting $\SU(2)$ holonomies with $\SL(2,\C)$ group elements.}
\label{fig:boostedholonomy}
\end{figure}

We have studied in details the case of a single loop insertion along the edge and mapped it on the insertion of a  boost generator in the $\sl(2,\C)$ Lie algebra. As we would like to consider multiple loop insertions, possibly towards a continuum of loop insertions as drawn on fig.\ref{fig:boostedholonomy}, it seems natural to imagine such a limit as described by a $\SL(2,\C)$ group element, somehow exponentiating the case of a single boost generator insertion. This would correspond to the $\SL(2,\C)$ holonomy along dressed edges as we describe for classical twisted geometries in section \ref{classical}.

Let us study the spin propagation defined by a $\SL(2,\C)$ group element  along a spin network edge.
Starting at the source of the edge with a given state $|j,m\ra$, a $\SL(2,\C)$ group element $G$ defines a probability amplitude for the state at the target, which depends on the choice of unitary representation $(n,\rho)$ on the edge:
\be
\cP_{j,m}^{(n,\rho)}[j',m';G]
\equiv
\left| 
\psi_{j,m}^{(n,\rho)}[j',m';G]
\right|^2
\,,\qquad
\psi_{j,m}^{(n,\rho)}[j',m';G]
\equiv
D^{(n,\rho)}_{(j,m),(j',m')}(G)
\,,
\ee
which is naturally normalized because we are working with a unitary representation of $\SL(2,\C)$:
\be
\sum_{j',m'} \cP_{j,m}^{(n,\rho)}[j',m';G]
=
\sum_{j',m'} D^{(n,\rho)}_{(j,m),(j',m')}(G)\,\overline{D^{(n,\rho)}_{(j,m),(j',m')}(G)}
=
\sum_{j',m'} D^{(n,\rho)}_{(j,m),(j',m')}(G)\,{D^{(n,\rho)}_{(j',m'),(j,m)}(G^{-1})}
=
1
\,.
\ee
If we cut the $\SL(2,\C)$ group element $G=e^{i\eta K}\in\SL(2,\C)$ into infinitesimal pieces, in the same fashion for the discretization of a path integral on an ordered exponentiation,
\beq
D^{(n,\rho)}_{(j,m),(j',m')}(e^{i\eta K})
&=&
\la (n,\rho)\, j,m |e^{i\eta K}| (n,\rho)\, j',m' \ra
\nn\\
&=&
%\sum_{\{j_{i},m_{i}\}_{i}}
%\la (n,\rho)\, j,m |(\id+\f iN\eta K)| (n,\rho)\, j_{1},m_{1} \ra
%\la (n,\rho)\, j_{1},m_{1} |(\id+\f iN\eta K)| (n,\rho)\, j_{2},m_{2} \ra
%\dots
%\la (n,\rho)\, j_{N-1},m_{N-1} |(\id+\f iN\eta K)| (n,\rho)\, j',m' \ra
\sum_{\{j_{k},m_{k}\}_{k}}
\prod_{k=0}^{N-1}
\Big{\la} (n,\rho)\, j_{k},m_{k} \Big{|}\,(\id+\f iN\eta K)\,\Big{|} (n,\rho)\, j_{k+1},m_{k+1} \Big{\ra}
\,,
\eeq
with $(j_{0},m_{0})=(j,m)$ and $(j_{N},m_{N})=(j',m')$, the matrix $D^{(n,\rho)}(G)$ becomes the product of a continuum of spin transitions induced by the action of boost $\sl(2,\C)$ generators.
More precisely, we get a superposition of trivial propagation by the identity $\id$ (which can be thought as a loop insertion with $J=0$) and boost generator insertions which can be identified as insertions of tadpoles with $J=1$ according to the results \eqref{boostshift} of the previous section \ref{CBsl2c} showing the equivalence between the action of boost generators and Clesch-Gordan coefficients at $J=1$.
This shows how that a $\SL(2,\C)$ holonomy along a spin network edge is equivalent to a continuum of insertions of tadpoles or little loops representing curvature excitations along the edge, as illustrated on fig.\ref{fig:boostedholonomy}.

\medskip

An arbitrary group element $G$ can be decomposed as $G=g_{2}\Lambda g_{1}$ with $g_{1},g_{2}$ in the $\SU(2)$ subgroup and $\Lambda$ a pure boost along $\sigma_{3}$. Since $\SU(2)$ group elements leave the spin $j$ invariant and do not create any spin shift, it is enough to investigate the action of a pure boost $\Lambda=\exp[\f\eta2\,\sigma_{3}]$ labeled by the boost  rapidity $\eta$. One can find the matrix elements of $\SL(2,\C)$ group elements in unitary representations, in e.g. \cite{AIHPA_1967__6_1_17_0}, expressed in terms of hypergeometric functions:
\beq
\cD^{(n,\rho)}_{(j,m),(j',m')}[\eta]
&=&
D^{(n,\rho)}_{(j,m),(j',m')}(e^{\f\eta2\,\sigma_{3}})
=
\la (n,\rho) j,m|e^{\f\eta2\,\sigma_{3}}|(n,\rho) j',m'\ra
\nn\\
&=&
\f{\delta_{m,m'}}{(j+j'+1)!}\,
\big{[}
(2j+1)(2j'+1)
\big{]}^{\f12}
\Delta_{j,m}\Delta_{j,n}\Delta_{j',m}\Delta_{j',n}
%\big{[}
%(j+m)!(j-m)!(j+n)!(j-n)!
%(j'+m')!(j'-m')!(j'+n)!(j'-n)!
%\big{]}^{\f12}
\nn\\
&&
\sum_{d,d'}
(-1)^{d+d'}
\f{(d+d'+m+n)!(j+j'-d-d'-m-n)!}{d!d'!(j-m-d)!(n+m+d)!(j-n-d)!(j'-m-d')!(n+m+d')!(j'-n-d')!}
\nn\\
&&
e^{-\eta\left(2d'+m+n+{i\rho}+1\right)}
\,F\Big{(}
j'+1+{i\rho},d+d'+m+n+1\,;j+j'+2;1-e^{-2\eta}
\Big{)}
\,,
\eeq
%\be
%\textrm{with}
%\qquad
%\Delta_{j,m}
%=
%\sqrt{(j+m)!(j-m)!}
%\,.
%\nn
%\ee
with the convention $\Delta_{j,m}=\sqrt{(j+m)!(j-m)!}$ for the pre-factors.
The range for the integer $d$ is  between 0 and the minimum of $(j-m)$ and $(j-n)$, and similarly for $d'$.

First, for $\eta=0$, it is straightforward to check that this gives the identity operator:
\be
\cD^{(n,\rho)}_{(j,m),(j',m')}[\eta]=\delta_{jj'}\delta_{mm'}
\,.
\ee
Then as the rapidity $\eta$ grows, we can get non-trivial transitions to spins $j'$ different from the initial spin $j$. We start by computing the expectation value for the target spin $j'$ by computing the action of the boost $\Lambda=\exp[\f\eta2\,\sigma_{3}]=\exp[i\eta\,K_{3}]$ on the $\SU(2)$-Casimir $\vJ^2$:
\be
\vJ^2\mapsto
\exp[-i\eta\,K_{3}]\vJ^2\exp[+i\eta\,K_{3}]
=
(\cosh\eta J_{1} -\sinh\eta K_{2})^2
+
(\cosh\eta J_{2} +\sinh\eta K_{1})^2
+J_{3}^2
\,.
\ee
Then using the explicit action of the $\sl(2,\C)$ generators in the $(n,\rho)$ unitary representation as given in appendix \ref{sl2C}, we obtain for a given source state $|j,m\ra$:
\beq
\sum_{j'}j'(j'+1)\cP_{j,m}^{(n,\rho)}[j',m;\eta]
&=&
\la(n,\rho)\,j,m| e^{-\eta\,K_{3}}\,\vJ^2\,e^{+\eta\,K_{3}} |(n,\rho)\,j,m\ra
\nn\\
&=&
j(j+1)+\sinh^2\eta
\Bigg{[}
j(j+1)\Big{[}
1+|\beta_{j}|^2+|\alpha_{j+1}|^2+|\alpha_{j}|^2
\Big{]}
\nn\\
&&\qquad\qquad\qquad\qquad
-m^2\Big{[}
1+|\beta_{j}|^2-|\alpha_{j+1}|^2-|\alpha_{j}|^2
\Big{]}
\nn\\
&&\qquad\qquad\qquad\qquad
+2j\,(|\alpha_{j+1}|^2-|\alpha_{j}|^2)
+2\,|\alpha_{j+1}|^2
\Bigg{]}
\label{exact}
\eeq
with the coefficients $\alpha_{j}$ and $\beta_{j}$ depending on the $(n,\rho)$  $\sl(2,C)$-representation label:
\be
|\alpha_{j}|^2=
\f14\f{(j^2-n^2)(j^2+\rho^2)}{j^2(j^2-\f14)}\,,
\qquad
\beta_{j}=\f{n\rho}{j(j+1)}
\ee
When $j$ is large, i.e. $j\gg1$, $j\gg n$, $j \gg \rho$, with $m$ scaling as $j$, the coefficients $|\alpha_{j}|^2$ goes to $\f14$ while the coefficients $\beta_{j}$ goes to 0, so that the whole expression does not depend at leading order in $j$ on the $\sl(2,\C)$ representation $(n,\rho)$ and simplifies to:
\be
\la j'(j'+1)\ra_{j,m}
%^{(n,\rho)}
\underset{j\gg 1}\sim
j(j+1)+\sinh^2\eta\Big{[}
\f32 j(j+1)-\f12 m^2
\Big{]}
\,.
\ee
For the maximal magnetic moment $m=j$, which represents a semi-classical coherent 3d vector of length $j$ in the $\hat{z}$-direction (see \cite{perelomov1986generalized} for the general theory of coherent states on Lie groups and \cite{Livine:2007vk,Livine:2007ya,Bianchi:2010gc} for the application of  $\SU(2)$ coherent states to spin networks and spinfoam), this gives a simple boost formula:
\be
\label{j'max}
\la j'(j'+1)\ra_{j,m=j}
\underset{j\gg 1}\sim
j(j+1)\,\cosh^2\eta
\,.
\ee
For simple representations with $n=0$, we can push the expansion further and compute the effect of $\rho$ on the expectation value:
\beq
\label{average-jt}
\la j'(j'+1)\ra_{j,m}^{(0,\rho)}
&\underset{j\gg 1}\sim&
j(j+1)+\sinh^2\eta\Big{[}
\f32 j(j+1)-\f12 m^2
+\f12(\rho^2+\f14)(1+\f{m^2}{j^2})+1
\Big{]}
\,,
%\qquad
\\
\la j'(j'+1)\ra_{j,m=j}^{(0,\rho)}
&\underset{j\gg 1}\sim&
j(j+1)+\sinh^2\eta\Big{[}
j(j+\f32)
+\rho^2+\f54
\Big{]}
\,.
\eeq
Although this expectation value gives us extremely interesting information about the average value of the target spin $j'$ after the boost $\Lambda$, we nevertheless miss the information about the spread of the probability distribution and possible oscillations.
We will complete this description of the boost action on the spin with numerical simulations.

\begin{figure}[h!]
%\centering
\begin{subfigure}[t]{.3\linewidth}
\includegraphics[width=50mm]{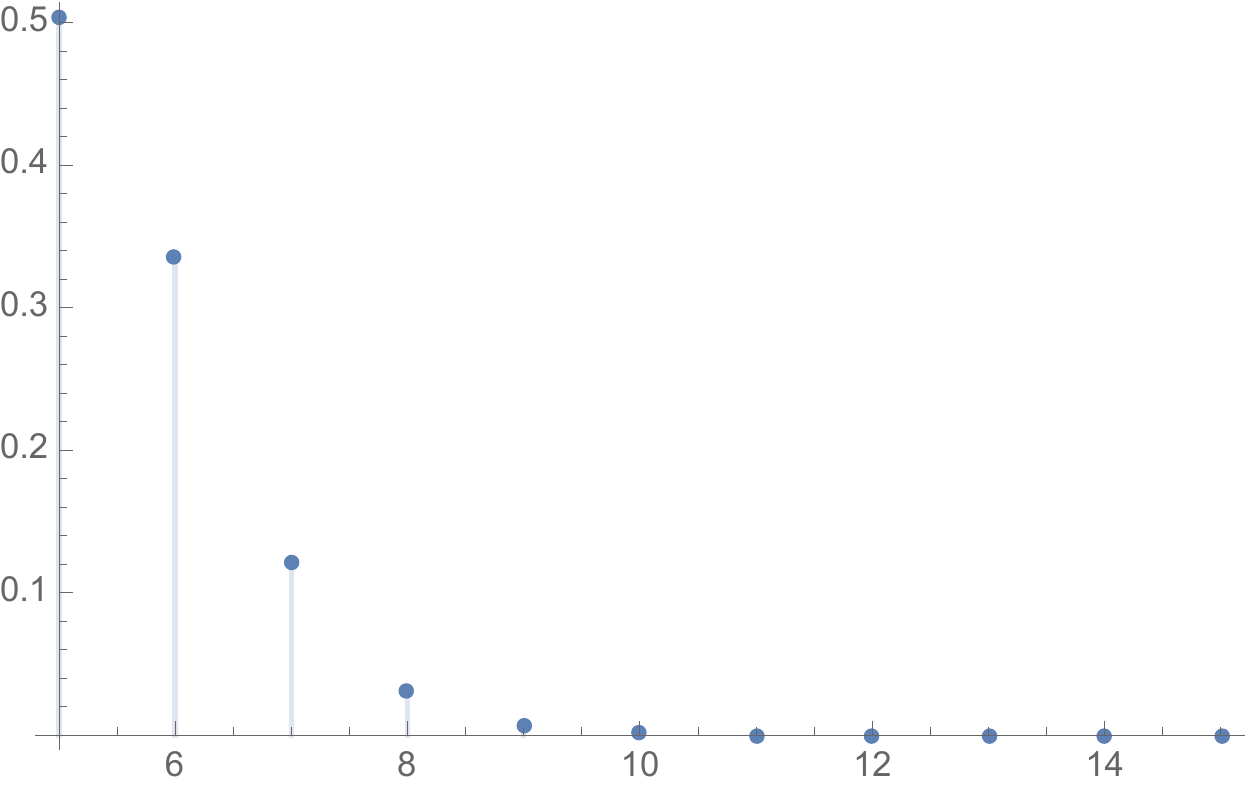}
\caption{Boost rapiditiy $\eta=0.5$, maximal target spin at $j'_{max}=5$ compared to $j\cosh\eta=5.64$, sum of displayed probabilities equal to 1.0000}
\label{fig:varyeta1}
\end{subfigure}
\hspace{2mm}
\begin{subfigure}[t]{.3\linewidth}
\includegraphics[width=50mm]{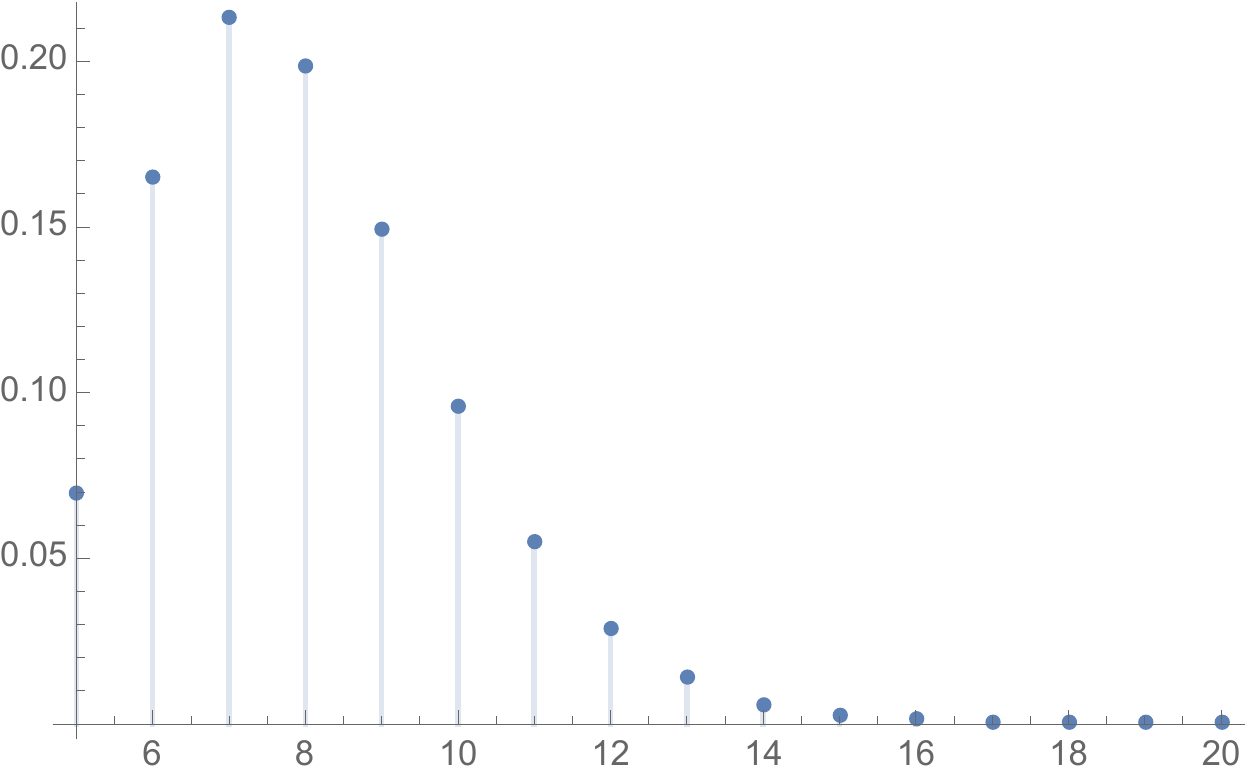}
\caption{Boost rapiditiy $\eta=1$, maximal target spin at $j'_{max}=7$ compared to $j\cosh\eta=7.72$, sum of displayed probabilities equal to 0.9999}
\label{fig:varyeta2}
\end{subfigure}
\hspace{2mm}
\begin{subfigure}[t]{.3\linewidth}
 \includegraphics[width=50mm]{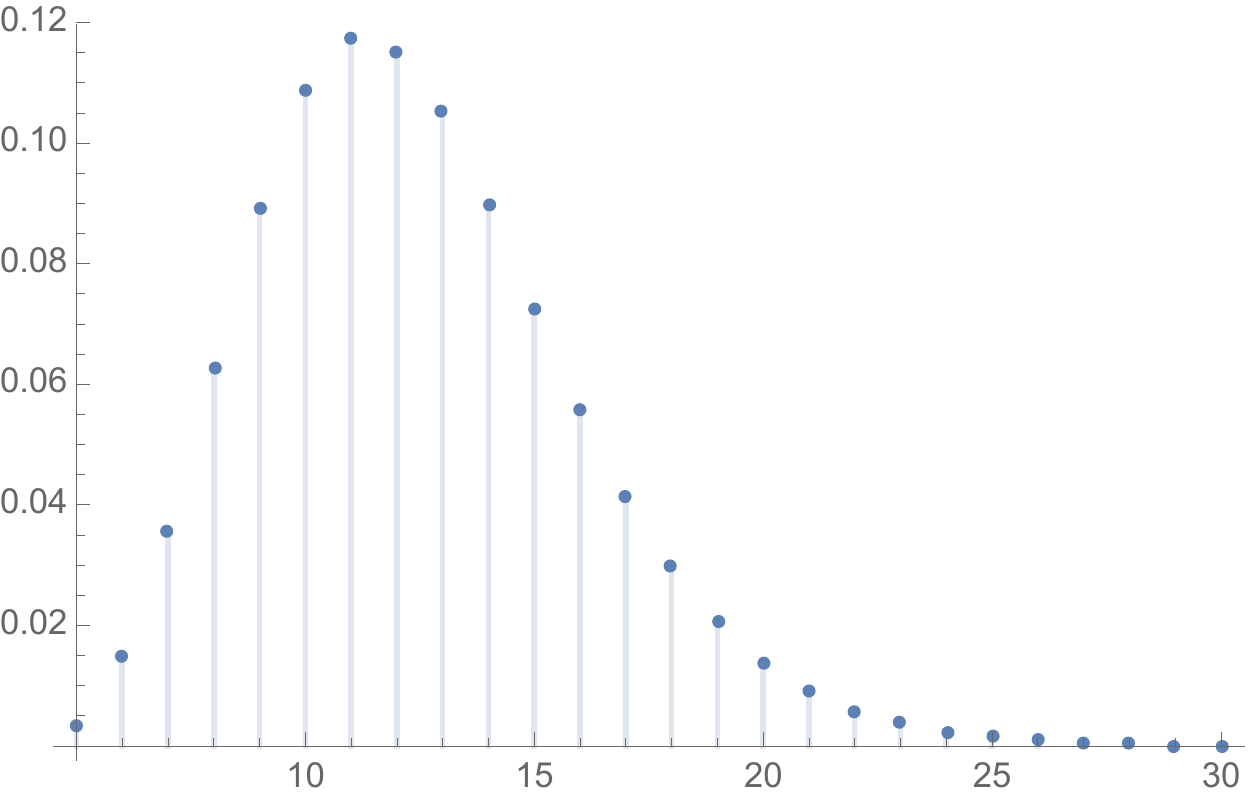}
\caption{Boost rapiditiy $\eta=1.5$, maximal target spin at $j'_{max}=11$ compared to $j\cosh\eta=11.76$, sum of displayed probabilities equal to 0.9998}
\label{fig:varyeta3}
\end{subfigure}

\begin{subfigure}[t]{.3\linewidth}
\includegraphics[width=50mm]{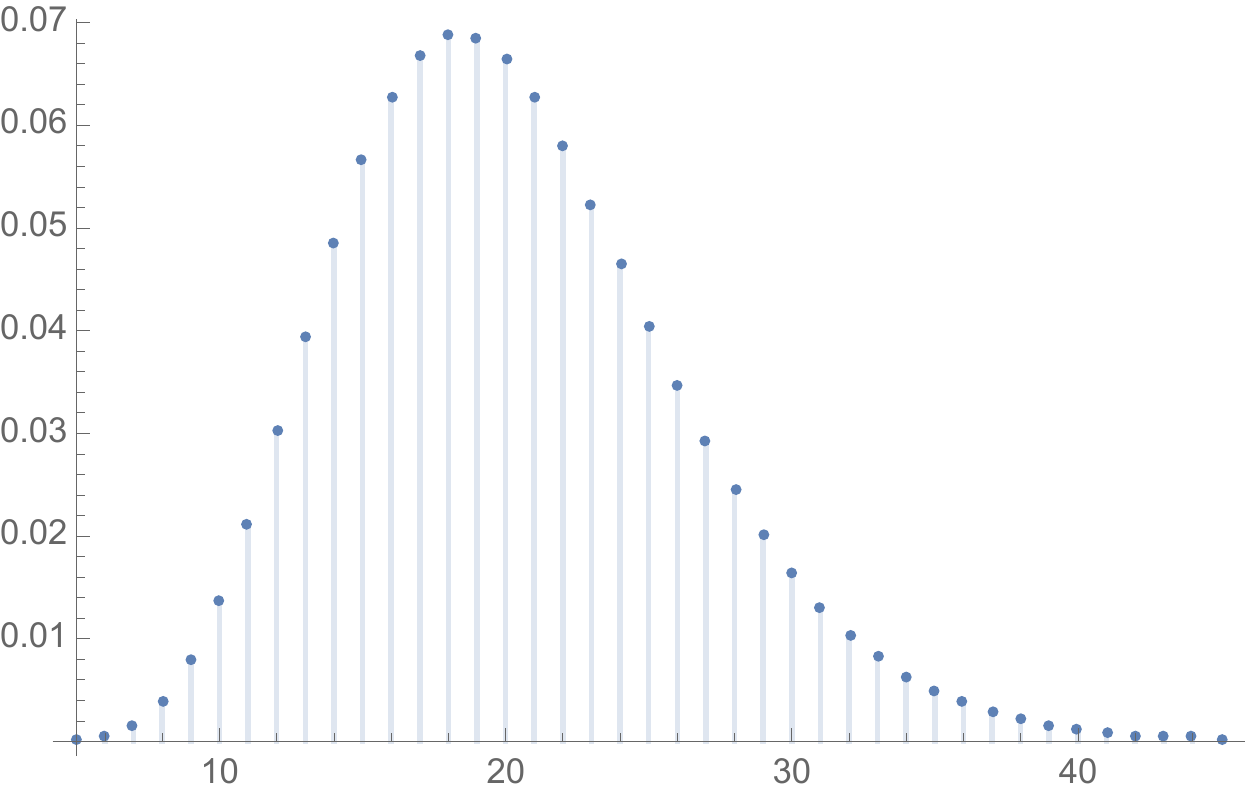}
\caption{Boost rapiditiy $\eta=2$, maximal target spin at $j'_{max}=18$ compared to $j\cosh\eta=18.81$, sum of displayed probabilities equal to 0.9993}
\label{fig:varyeta4}
\end{subfigure}
\hspace{2mm}
\begin{subfigure}[t]{.3\linewidth}
\includegraphics[width=50mm]{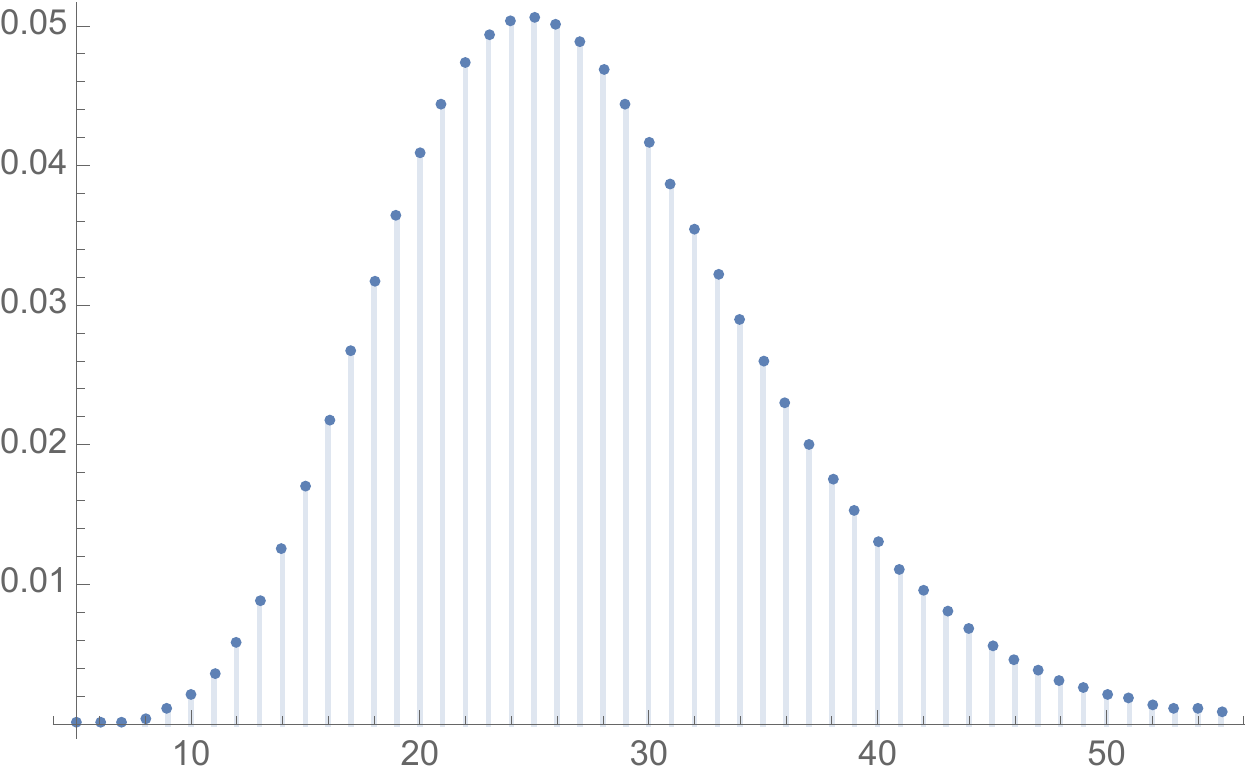}
\caption{Boost rapiditiy $\eta=2.3$, maximal target spin at $j'_{max}=25$ compared to $j\cosh\eta=25.19$, sum of displayed probabilities equal to 0.9974}
\label{fig:varyeta5}
\end{subfigure}
\hspace{2mm}
\begin{subfigure}[t]{.3\linewidth}
 \includegraphics[width=50mm]{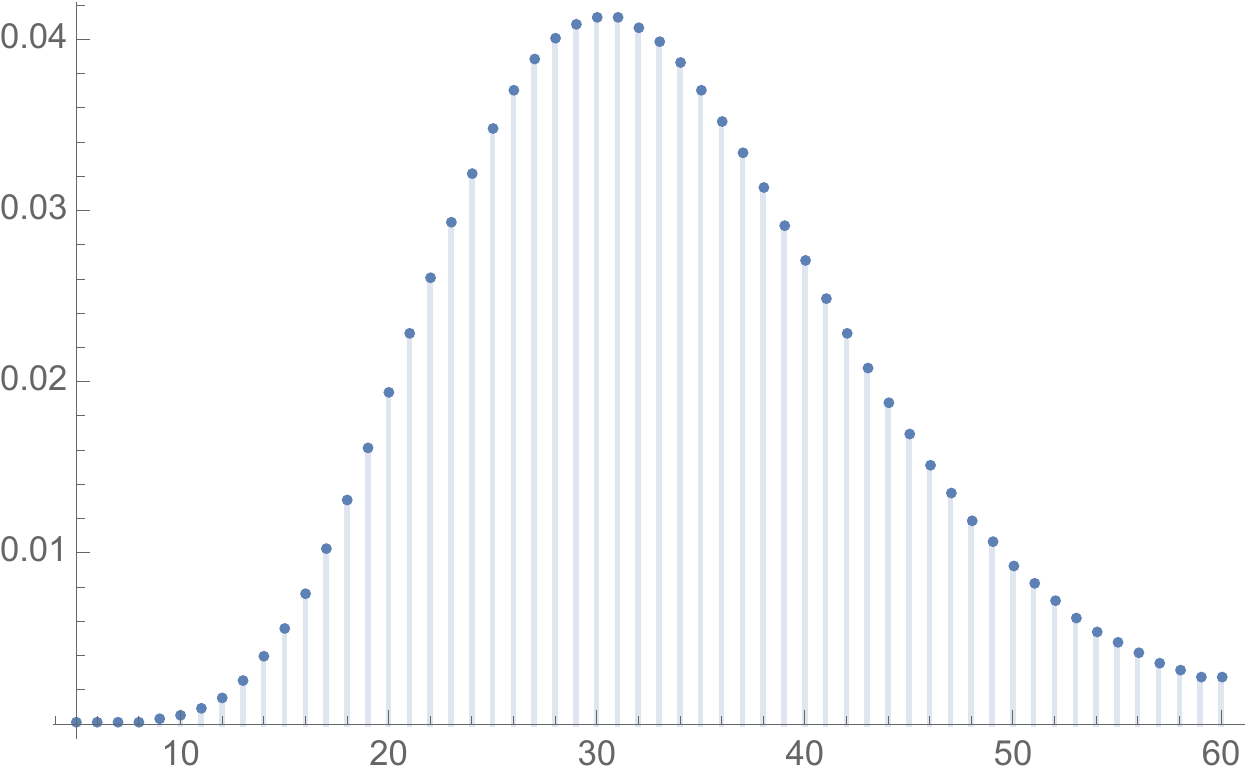}
\caption{Boost rapiditiy $\eta=2.5$, maximal target spin at $j'_{max}=30$ compared to $j\cosh\eta=30.66$, sum of displayed probabilities equal to 0.9904}
\label{fig:varyeta6}
\end{subfigure}

\caption{Plots of the probability distribution $\cP_{j,m}^{(n,\rho)}[j',m;\eta]$ in terms of the target spin $j'$ for trivial $\sl(2,\C)$ representation labels $(n,\rho)=(0,0)$, a fixed source spin $j=5$ and maximal magnetic moment $m=j=5$, for boost rapidities $\eta$ ranging from 0 to 2.5. We compare the value of the target spin with maximal probability $j'_{max}$ with the leading order of the expectation value $j'\sim j\cosh\eta$ obtained in \eqref{j'max}.}
\label{fig:varyeta}
\end{figure}
\begin{figure}[h!]
%\centering

\begin{subfigure}[t]{.48\linewidth}
 \includegraphics[width=60mm]{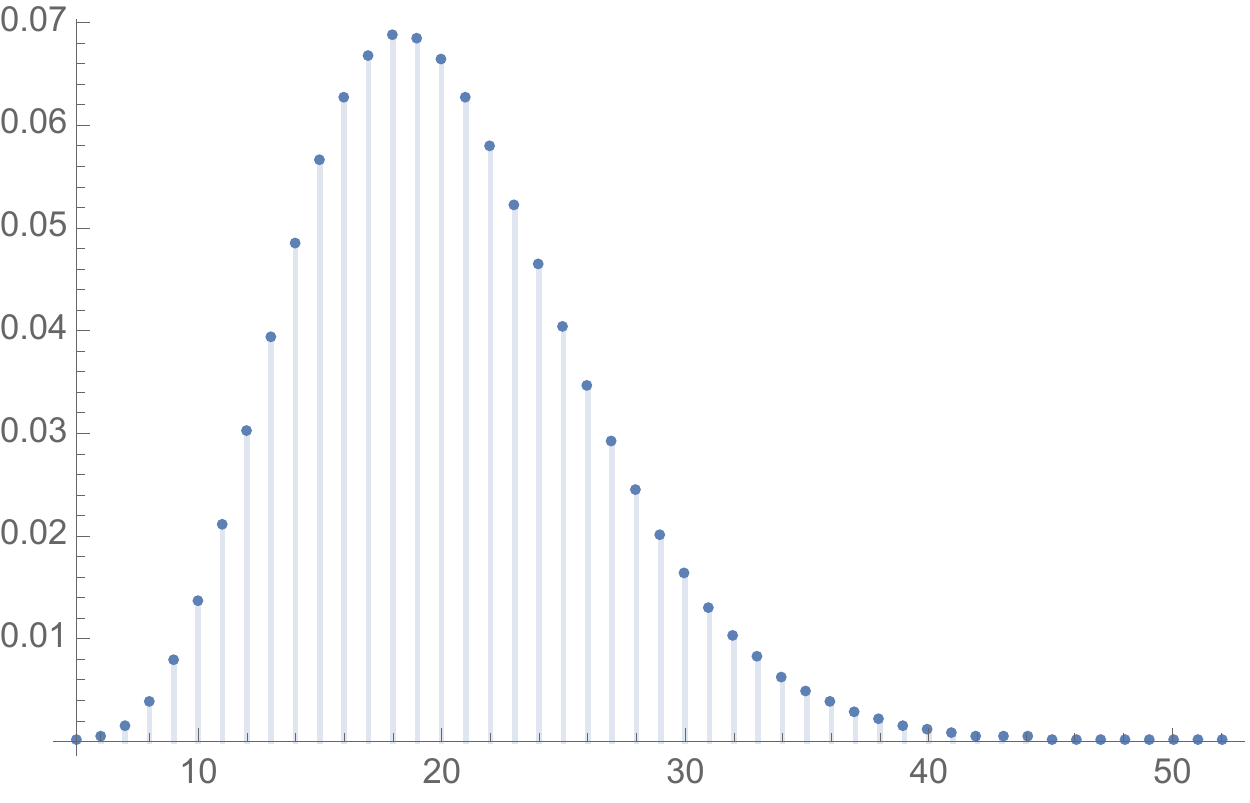}
\caption{$\sl(2,\C)$ representation labels $(n,\rho)=(0,0)$, sum of displayed probabilities equal to 1.000, expectation values $\la j'(j'+1) \ra_{num}=467.0$ and $\la j'(j'+1) \ra_{ana}=467.1$}
\label{fig:varyrho0}
\end{subfigure}
\hspace{2mm}
\begin{subfigure}[t]{.48\linewidth}
 \includegraphics[width=60mm]{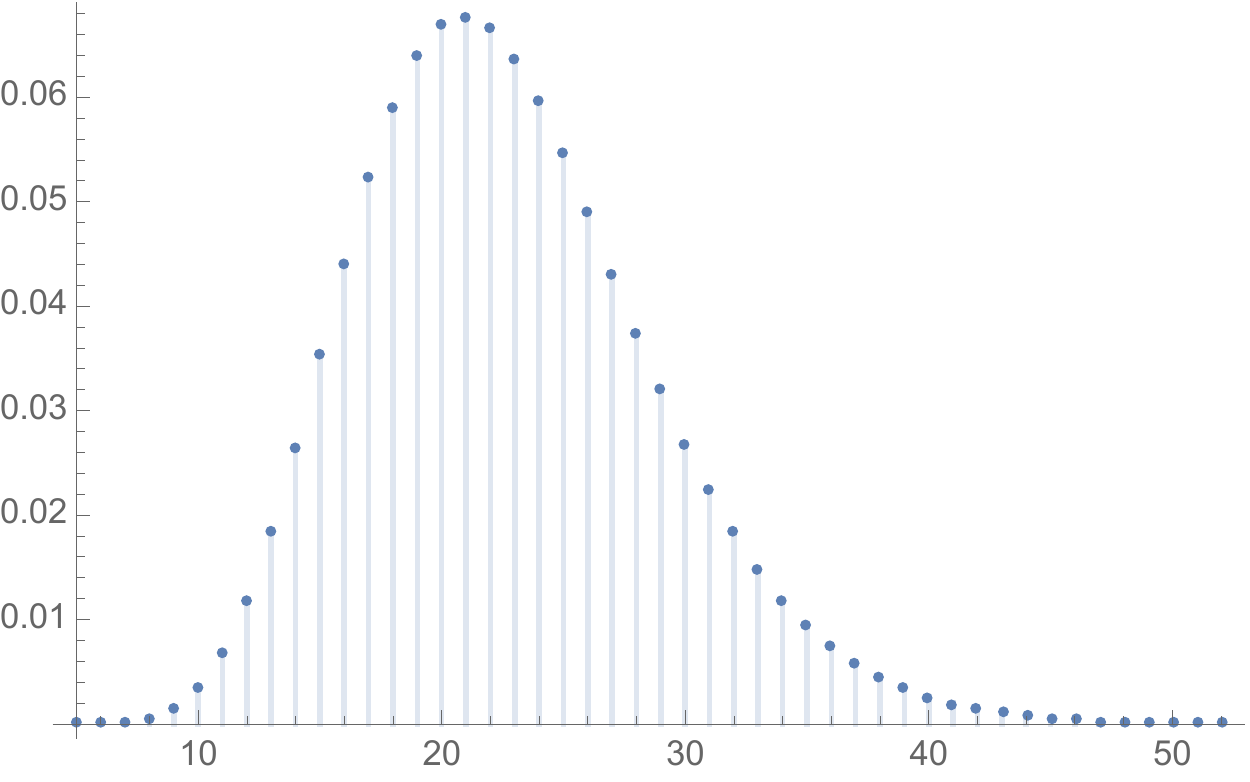}
\caption{$\sl(2,\C)$ representation labels $(n,\rho)=(0,3)$, sum of displayed probabilities equal to 1.000, expectation values $\la j'(j'+1) \ra_{num}=575.8$ and $\la j'(j'+1) \ra_{ana}=576.4$}
\label{fig:varyrho3}
\end{subfigure}

\begin{subfigure}[t]{.48\linewidth}
 \includegraphics[width=60mm]{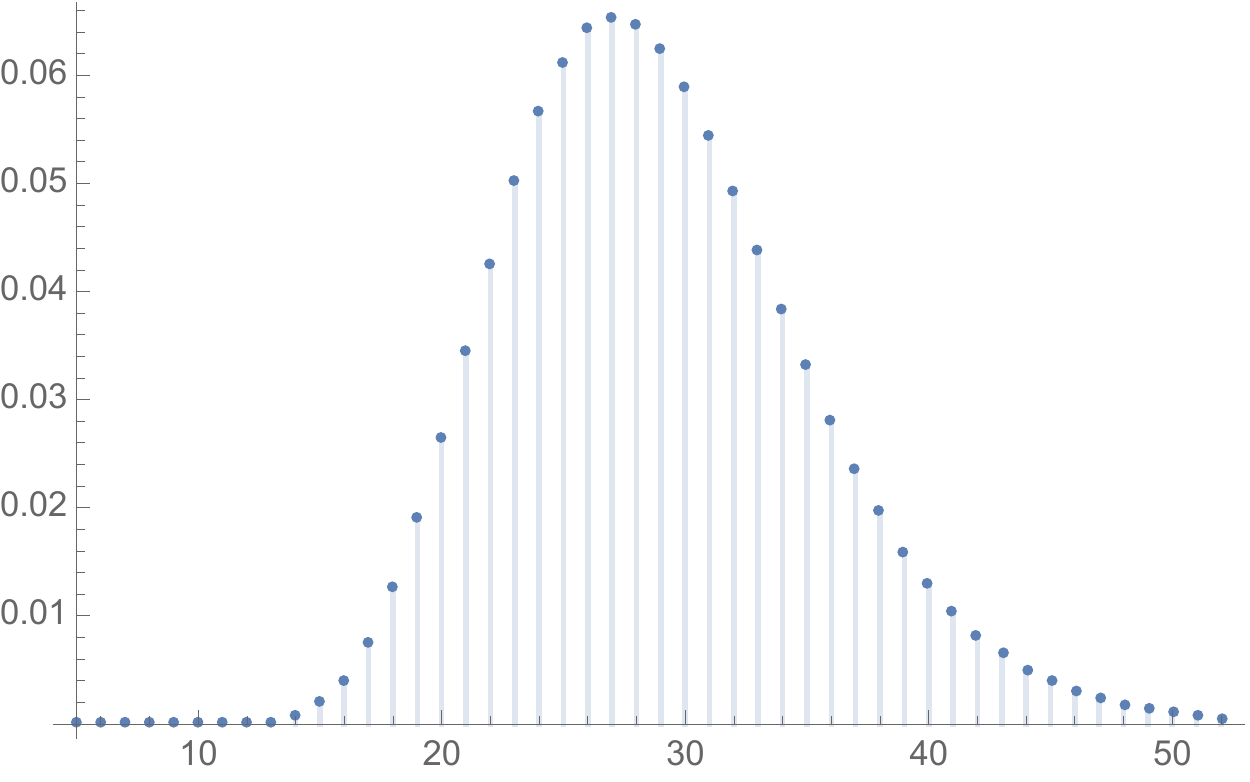}
\caption{$\sl(2,\C)$ representation labels $(n,\rho)=(0,6)$, sum of displayed probabilities equal to 0.999, expectation values $\la j'(j'+1) \ra_{num}=900.0$ and $\la j'(j'+1) \ra_{ana}=904.2$}
\label{fig:varyrho6}
\end{subfigure}
\hspace{2mm}
\begin{subfigure}[t]{.48\linewidth}
 \includegraphics[width=60mm]{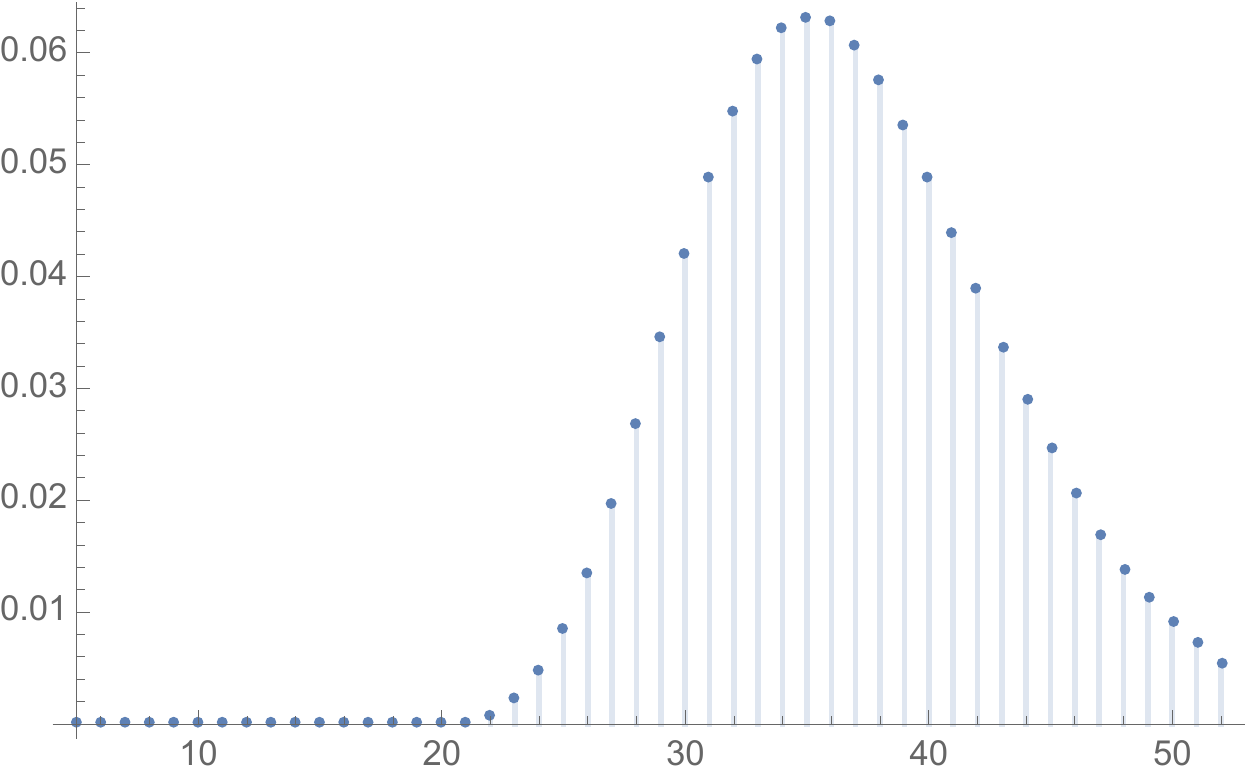}
\caption{$\sl(2,\C)$ representation labels $(n,\rho)=(0,9)$, sum of displayed probabilities equal to 0.981, expectation values $\la j'(j'+1) \ra_{num}=1389.7$ and $\la j'(j'+1) \ra_{ana}=1450.6$ (slight decrease in numerical precision)}
\label{fig:varyrho9}
\end{subfigure}

\caption{Plots of the probability distribution $\cP_{j,m}^{(n,\rho)}[j',m;\eta]$ in terms of the target spin $j'$, for a fixed source state $j=m=5$, boost rapidity $\eta=2$ and $\sl(2,\C)$ spin $n=0$, as we vary $\rho$ from 0 to 9. We see that  the peak for the target spin $j'_{max}$ increases with as $\rho$ grows. We can compare the numerical value for the expectation value $\la j'(j'+1) \ra_{num}$ to its analytical prediction $\la j'(j'+1) \ra_{ana}$ given by \eqref{exact}. The numerical value is consistently smaller than the analytical computation since we have truncated the numerical probability distribution to $j'\le 52$.}
\label{fig:varyrho}
\end{figure}
\begin{figure}
%\centering
\begin{subfigure}[t]{.3\linewidth}
\includegraphics[width=50mm]{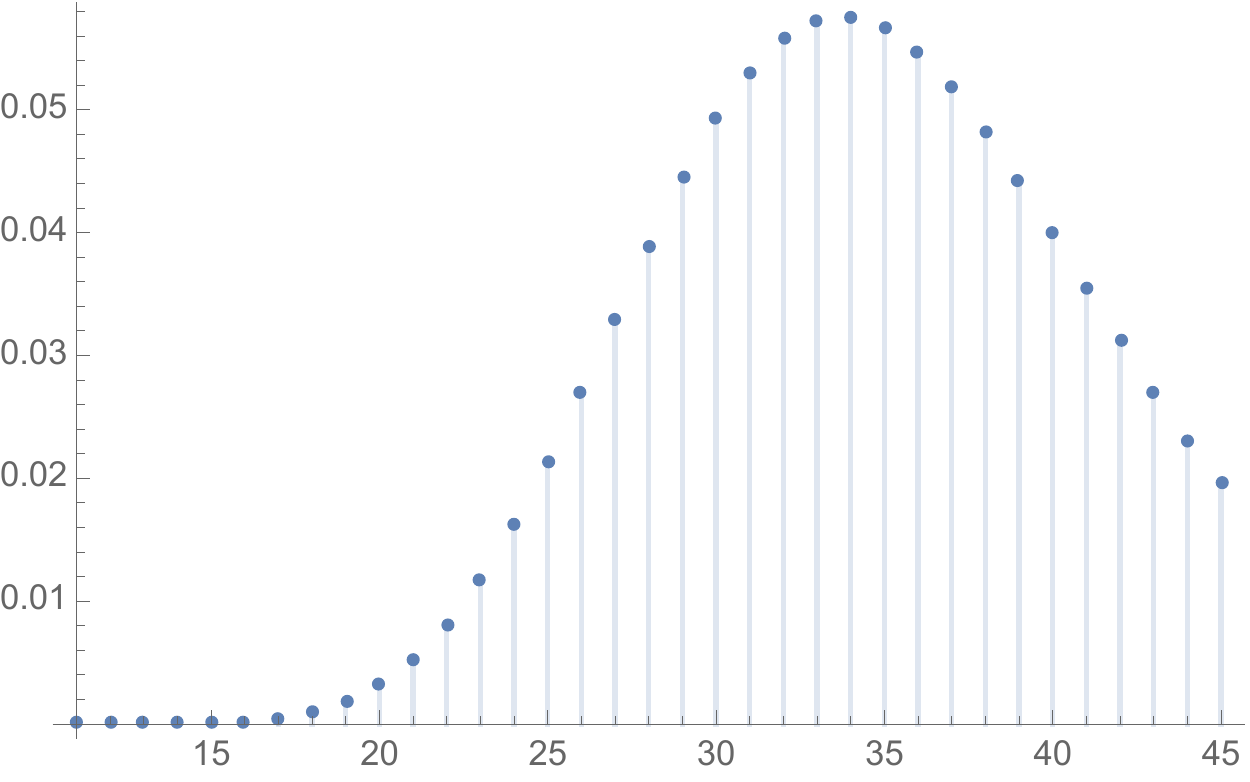}
\caption{$n=0\le j=11$}
\label{fig:varyn0}
\end{subfigure}
\hspace{2mm}
\begin{subfigure}[t]{.3\linewidth}
\includegraphics[width=50mm]{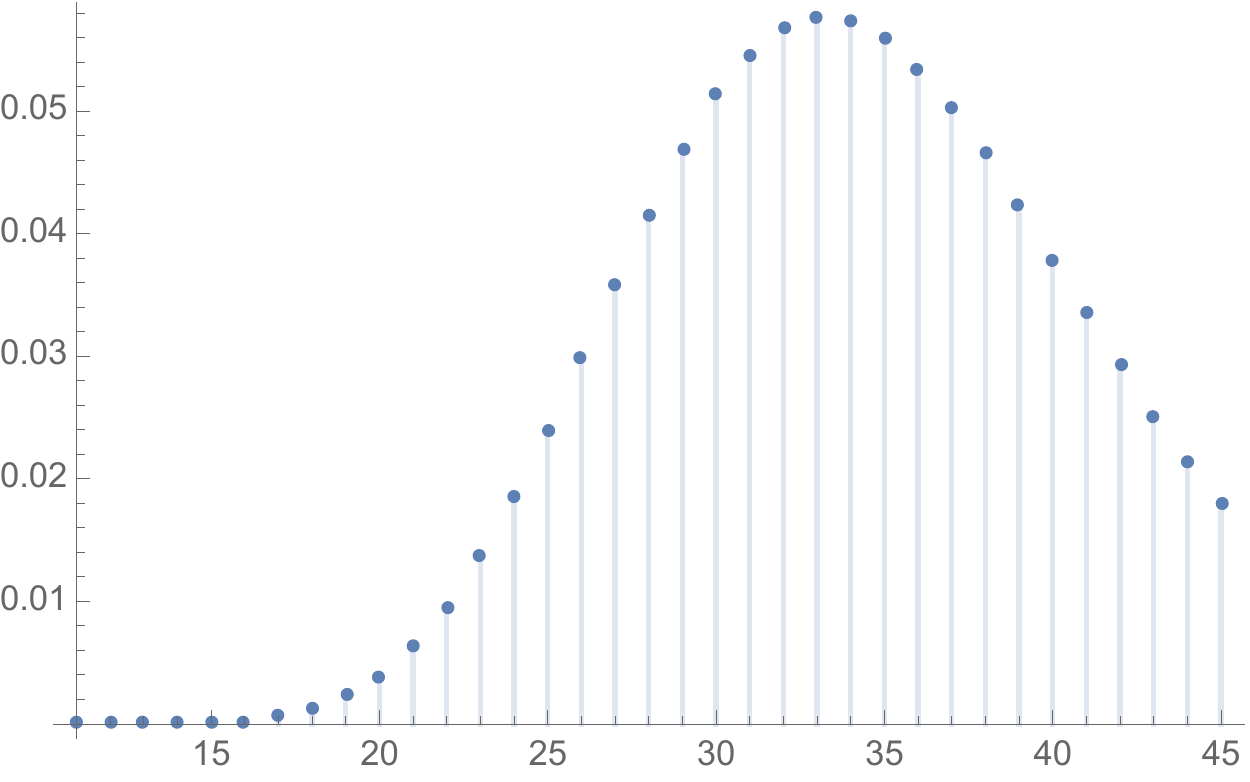}
\caption{$n=2\le j=11$}
\label{fig:varyn2}
\end{subfigure}
\hspace{2mm}
\begin{subfigure}[t]{.3\linewidth}
\includegraphics[width=50mm]{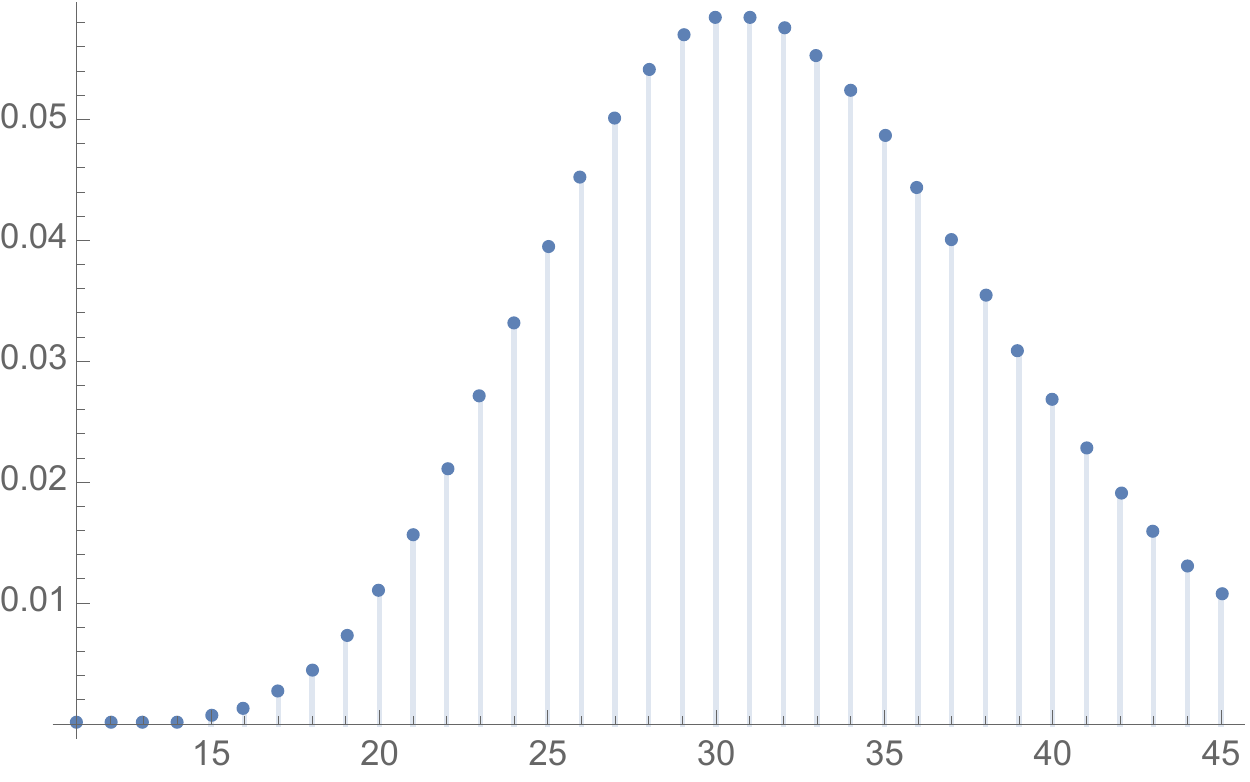}
\caption{$n=5\le j=11$}
\label{fig:varyn5}
\end{subfigure}

\begin{subfigure}[t]{.3\linewidth}
\includegraphics[width=50mm]{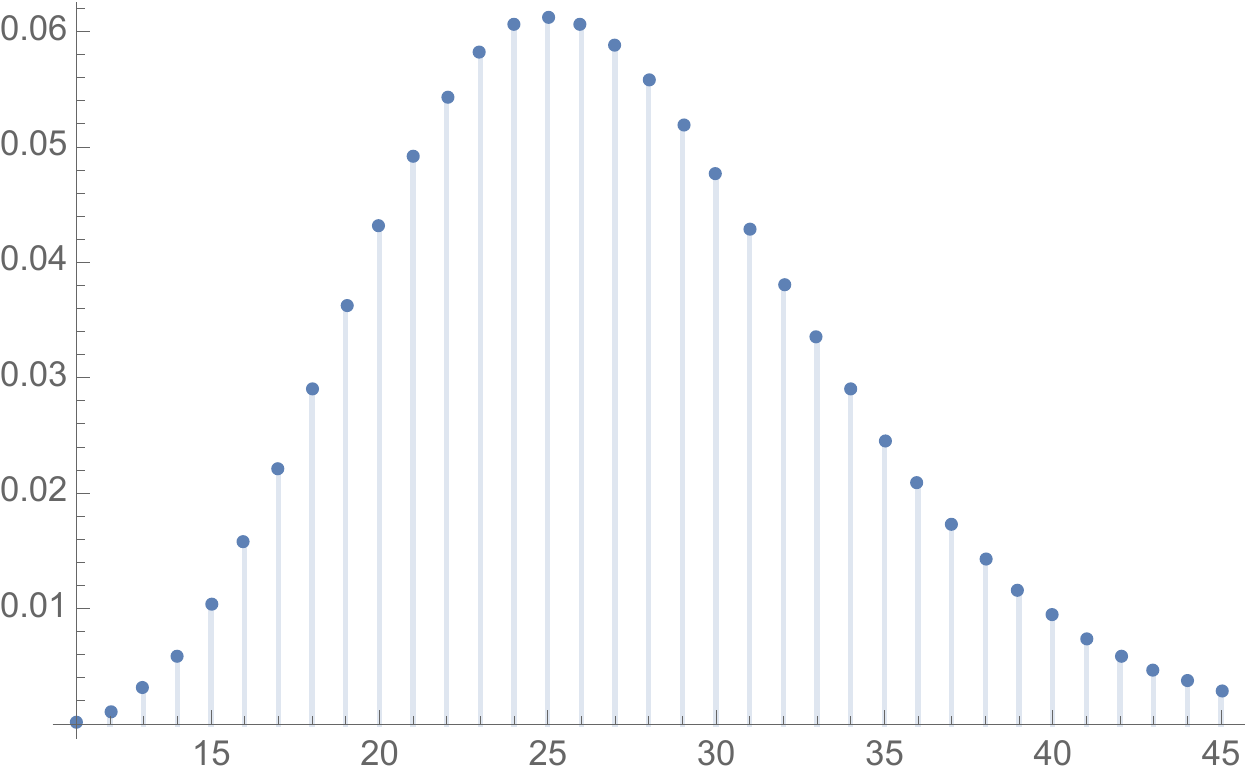}
\caption{$n=8\le j=11$}
\label{fig:varyn8}
\end{subfigure}
\hspace{2mm}
\begin{subfigure}[t]{.3\linewidth}
\includegraphics[width=50mm]{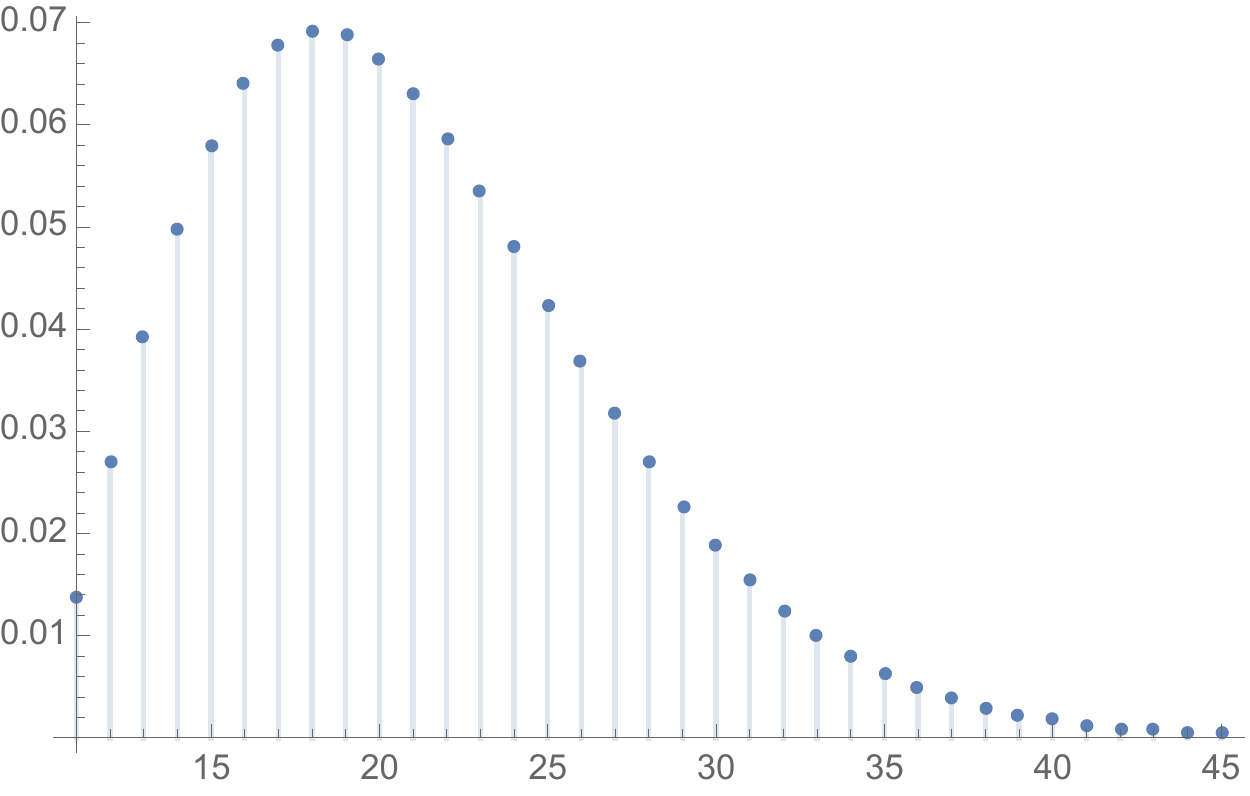}
\caption{$n=10\le j=11$}
\label{fig:varyn10}
\end{subfigure}
\hspace{2mm}
\begin{subfigure}[t]{.3\linewidth}
 \includegraphics[width=50mm]{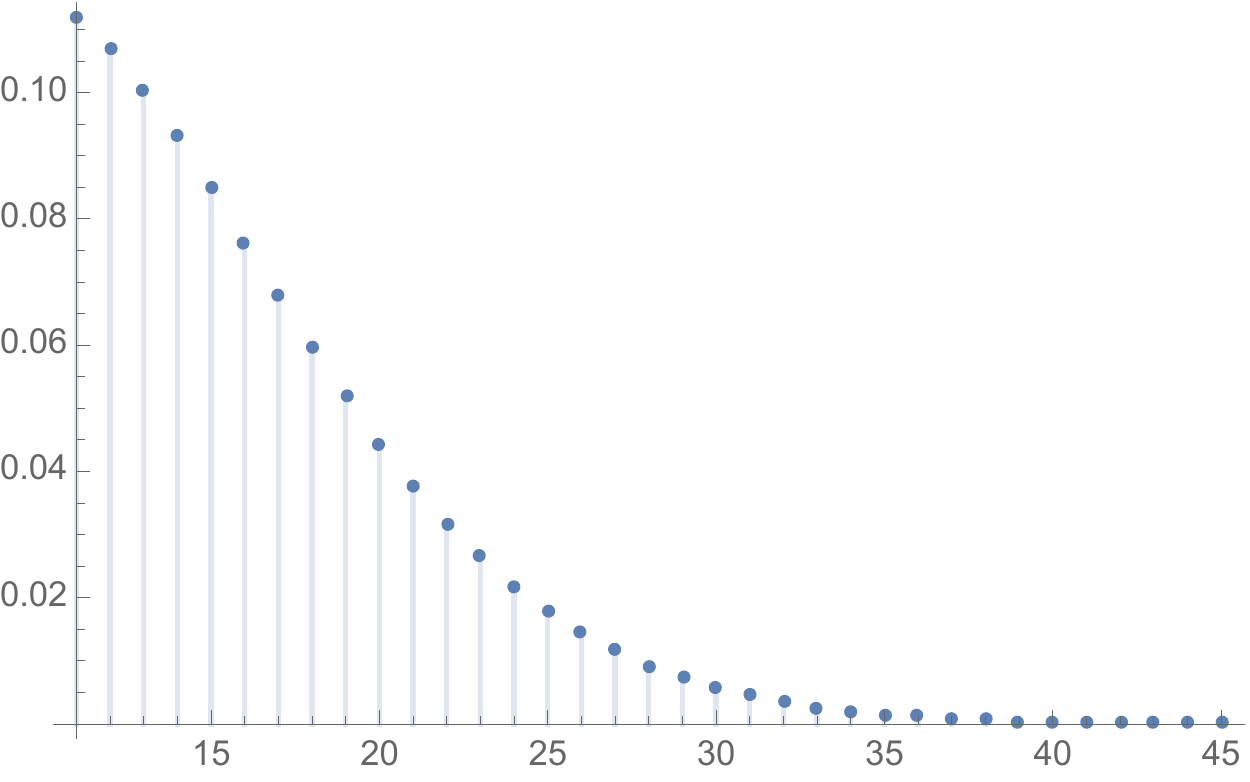}
\caption{$n=11=j$}
\label{fig:varyn11}
\end{subfigure}

\caption{Plots of the probability distribution $\cP_{j,m}^{(n,\rho)}[j',m;\eta]$ for $j=m=11$ and boost rapidity $\eta=1.8$ in terms of the target spin $j'$  as we vary the $\sl(2,\C)$ representation label $n$ from 0 to its maximal value $j=11$ while keeping $\rho=0$ fixed. As compared to the behavior when varying $\rho$, the peak is displaced toward $j'=j$ as we increase $n$ from 0 to $j$.}
\label{fig:varyn}
\end{figure}
\begin{figure}
%\centering

\begin{subfigure}[t]{.23\linewidth}
\includegraphics[width=40mm]{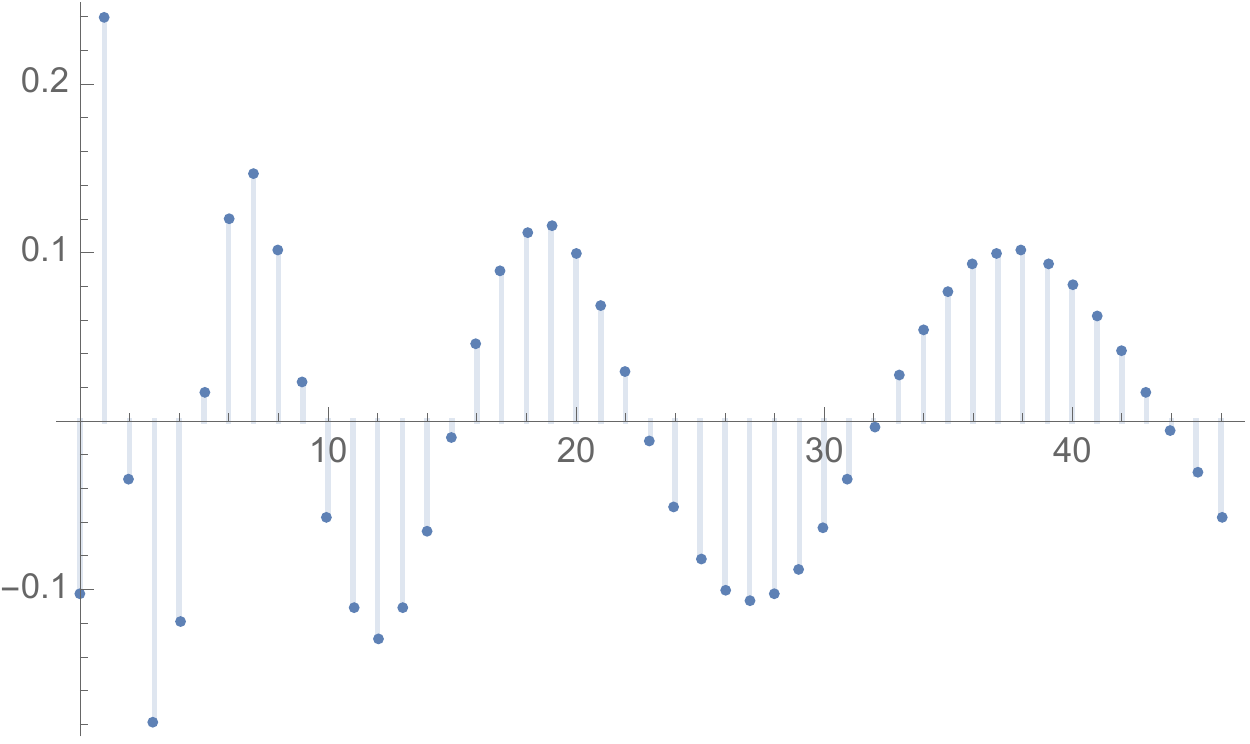}
\caption{$m=0< j=11$}
\label{fig:varym0}
\end{subfigure}
\hspace{1mm}
\begin{subfigure}[t]{.23\linewidth}
\includegraphics[width=40mm]{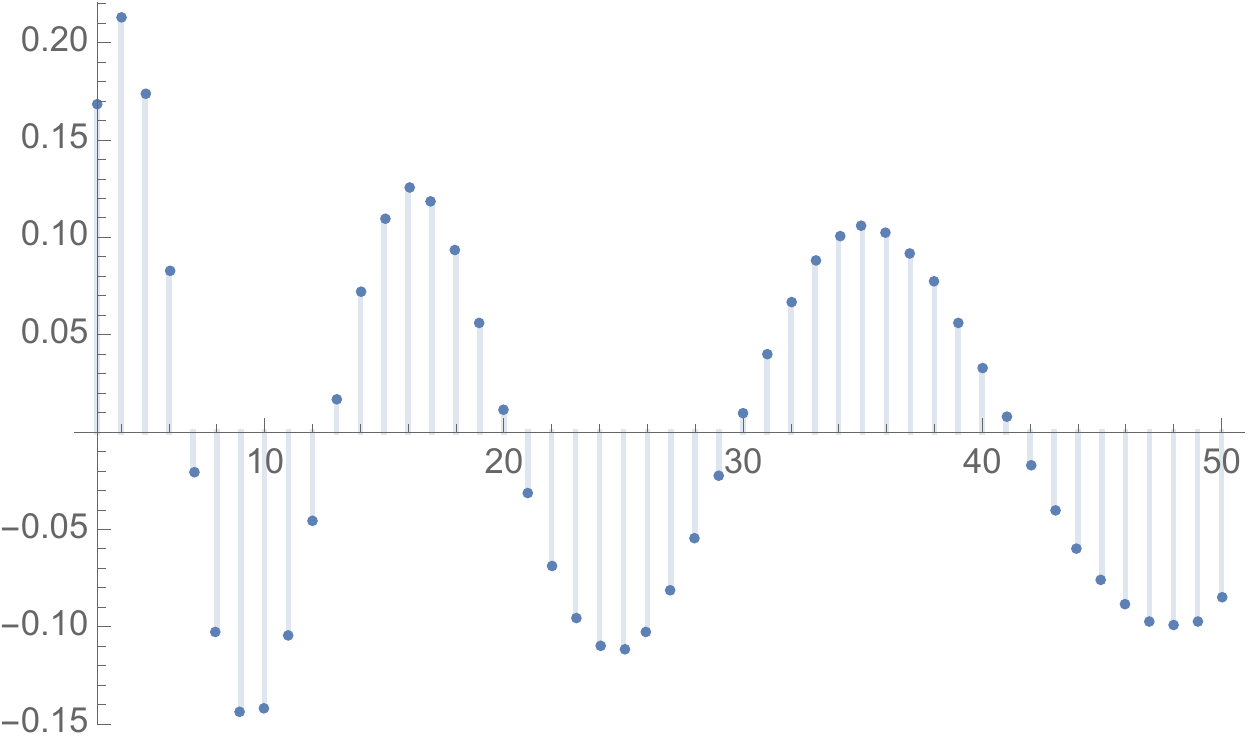}
\caption{$m=3< j=11$}
\label{fig:varym3}
\end{subfigure}
\hspace{1mm}
\begin{subfigure}[t]{.23\linewidth}
\includegraphics[width=40mm]{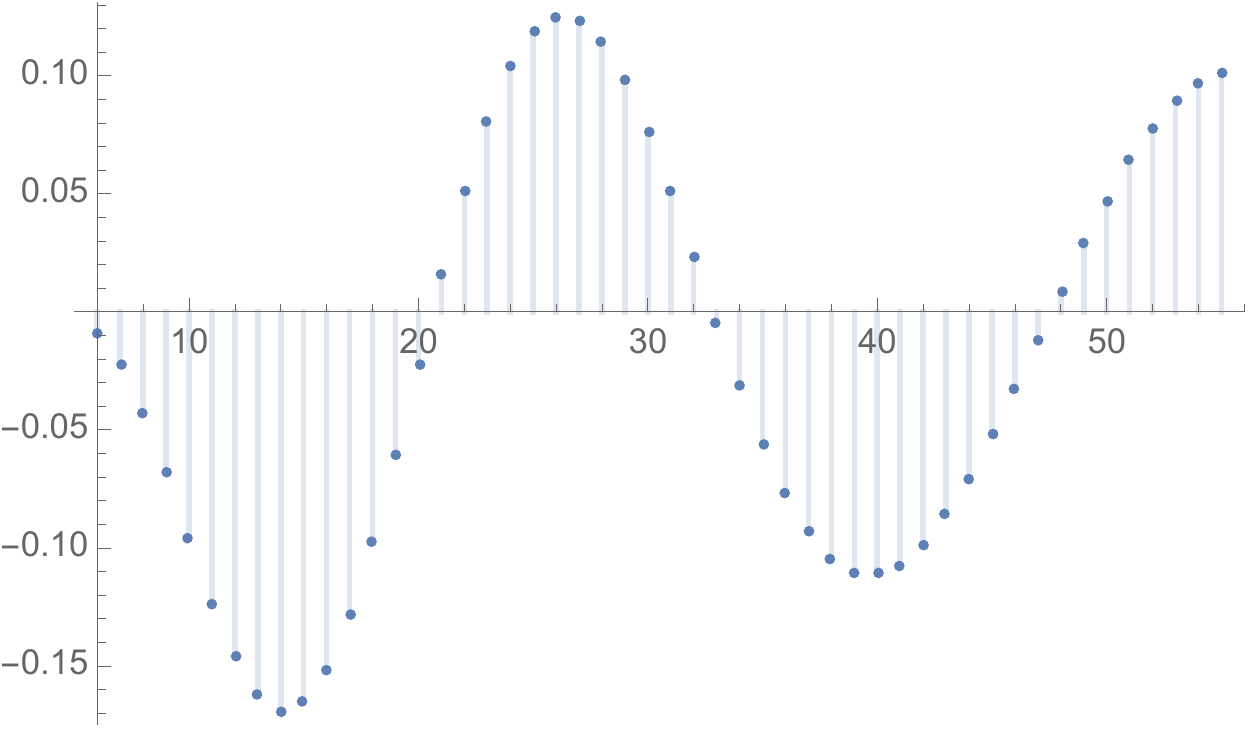}
\caption{$m=6< j=11$}
\label{fig:varym6}
\end{subfigure}
\hspace{1mm}
\begin{subfigure}[t]{.23\linewidth}
\includegraphics[width=40mm]{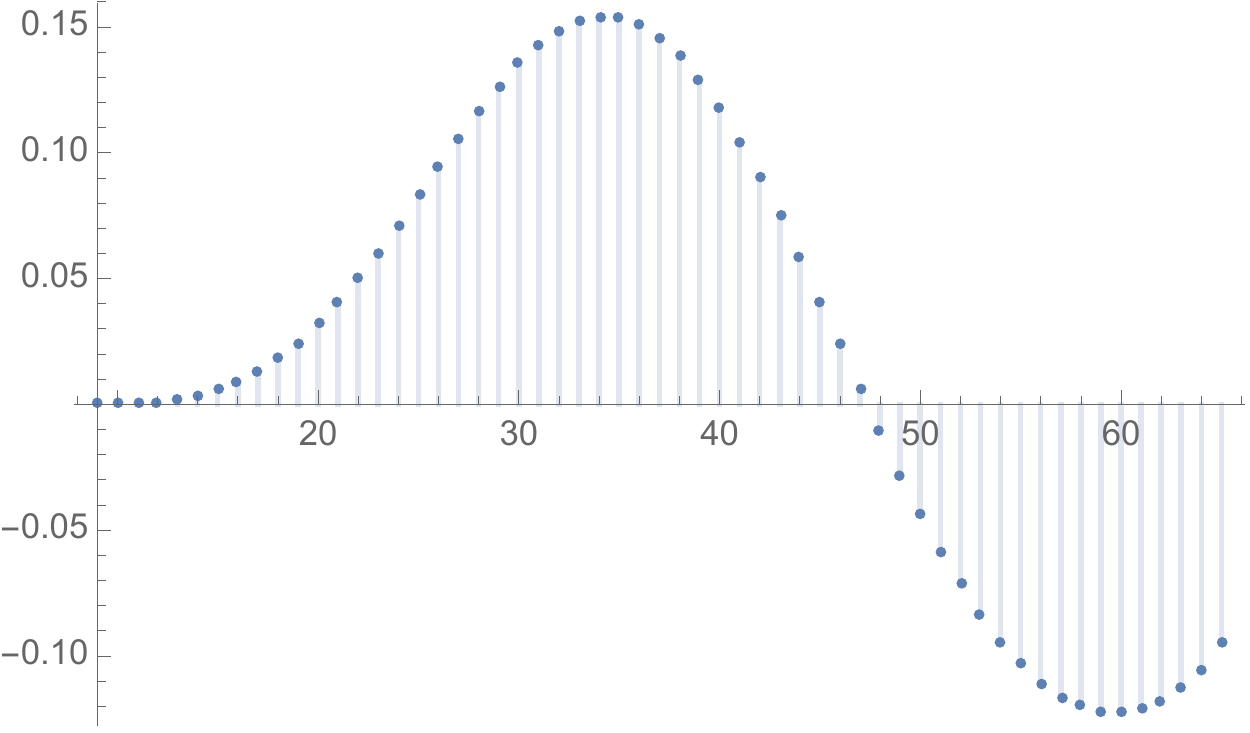}
\caption{$m=9< j=11$}
\label{fig:varym9}
\end{subfigure}

\caption{Plots of the probability amplitude $D_{j,m}^{(0,0)}[j',m;\eta]$ in terms of the target spin $j'$, for a fixed source spin $j=5$ and boost rapidity $\eta=2.5$ as we vary the magnetic moment $m$: compared to the $m=j$ case where the probability distribution has a single peak, it acquires oscillations as soon as $|m|<j$.}
\label{fig:varym}
\end{figure}

Let us start by looking at the case for the maximal allowed value for the magnetic moment label $m=j$, in which case the expression for the boost matrix simplifies:
\beq
\cD^{(n,\rho)}_{j,j'}[\eta]
&=&
\f{\sqrt{2j'+1}}{j+j'+1}\sqrt{\f{(2j+1)!(j'-j)!}{(j'+j)!}}\,\f{j'!}{j!}
\nn\\
&&\sum_{d=0}^{j'-j}\f1{d!(j'-j-d)!}
e^{-\eta\,(2d+j+i\rho+1)}
\,
F(j'+i\rho+1,d+j+1;j'+j+2;1-e^{-2\eta})
\,.
\eeq
First of all the target spin $j'$ is necessary larger or equal to $m'=m$ and thus $j'\ge j$. Then we see that the probability distribution\footnotemark{} $|\cD^{(n,\rho)}_{j,j'}[\eta]|^2$ for the target spin $j'$ is loosely peaked on an optimal value depending on the source spin $j$, on the boost rapidity $\eta$ and on the $\sl(2,\C)$ representation labels, as showed on  figures \ref{fig:varyeta}  \ref{fig:varyrho} \ref{fig:varyn} \ref{fig:varym}.
\footnotetext{
Focusing on the probability distribution $|\cD^{(n,\rho)}_{j,j'}[\eta]|^2$ for the target spin $j'$, we overlook the phase of the probability amplitude given by the boost matrix element $\cD^{(n,\rho)}_{j,j'}[\eta]$ itself. For instance, for a non-vanishing $\sl(2,\C)$ representation label $\rho>0$, these matrix elements are complex and their phase oscillates in terms of the boost rapidity $\eta$ with a frequency given by $\rho$. These correspond to variations in the conjugate variable to the spin living on the spin network edge, that is the twist angle (which is related to the extrinsic curvature) \cite{Freidel:2010aq,Anza:2014tea}.
}
More precisely, when the boost rapidity $\eta$ vanishes, the distribution is sharply peaked on the source spin $j'=j$. Then as the rapidity $\eta$ grows, the spin gets boosted and the optimal target spin increases, as showed on the plots\footnotemark{} of fig.\ref{fig:varyeta}. We can check that the shift of the optimal target spin fits with the analytical calculation of the expectation value done above in \eqref{average-jt}.
\footnotetext{We have plotted the probability distribution computed by brute force on Mathematica from the sum over the evaluation of the hypergeometric function. For more precise results, a more systematic approach could be to derive recursion relations satisfied by the hypergeometric function and sums involved in the matrix element $\cD^{(n,\rho)}_{(j,m),(j',m')}[\eta]$.}
The $\sl(2,\C)$ representation labels $(n,\rho)$ shifts the optimal target spin and the probability distribution towards higher values as $n$ or $\rho$ grow, as illustrated on figures \ref{fig:varyrho} and \ref{fig:varyn}.
Finally, we can explore other values for $m$ besides its maximal allowed value. It turns out that taking $m\ne \pm j$ leads to oscillations\footnotemark, on we can see on the multiples plots of fig. \ref{fig:varym}.
\footnotetext{
It could be interesting to switch to $\SU(2)$ coherent states instead of the $|j,m\ra$ basis, as used for coherent spin network states \cite{Livine:2007vk,Dupuis:2011fz} and for construction and semi-classical study of spinfoam models \cite{Livine:2007vk,Barrett:2009gg,Barrett:2009as}. They are $\SU(2)$ rotations of the highest  weight state $|j,m=j\ra$. We expect that boosting such states would not lead to oscillations but would keep a steady wave packets as for the $m=j$ case. This nevertheless remains to be studied. Moreover, we could also revisit the present discussion using $\SL(2,\C)$ coherent states, as used in Lorentzian spinfoam models \cite{Freidel:2007py,Barrett:2009mw,}.
}
This illustrates the two simple scenarii of spin propagation discussed in the previous section \ref{spinwave},  of a spin shift and of spin waves induced by the action of a $\SL(2,\C)$ holonomy on the $\SU(2)$ states on a spin network.

Getting a more precise picture of the spread of the probability distribution would involve gaining a better understanding of the asymptotics of  hypergeometric function  and controlling the sum over $d$ and $d'$. For instance, for large $j$, assuming that $j'$ and $m$ scale linearly with $j$ so that the ratios $j'/j\ge 1$ and $|m/j| \le 1$ are fixed, we can look at the asymptotic behavior of the hypergeometric function in terms of the ratios $x=d/j$ and $x'=d'/j$, as worked out e.g. as a steepest descent approximation of the integral form of the hypergeometric function in \cite{hyper1,hyper2}, and then compute the sum as a double Riemann integral. This seems highly technical and a qualitative description of the probability distribution is enough for the purpose of the present paper.

%%%%%%%%
\section*{Conclusion \& Outlook}
%%%%%%%%

This work has focused on the coarse-graining of spin network states in loop quantum gravity and their counterpart in classical geometry, twisted geometries. The goal was to study more precisely the coarse-graining of spin network edges and the algebraic data that they carry.
We have presented how loop insertions along the links of twisted geometries and spin networks represent curvature excitations and showed that their main effect is to relax the area-matching or spin-matching constraints along those links. Due to those possible curvature excitations, the spins at the source and target of a spin network edge can be different and the spin can further fluctuate along the edge. This can be interpreted as a non-trivial area propagator along spin network edges. Similarly to the renormalization of the propagator or 2-point function in quantum field theory, deriving the coarse-graining flow of this area propagator could be key to better understanding the coarse-graining and renormalization of spin networks and their quantum geometry.

We further showed how loop insertions along an edge is equivalent to the insertion of $\sl(2,\C)$ boost generators and that  the non-trivial propagation of the area and spin along an edge can be understood in the continuum limit as a $\SL(2,\C)$ transport along the edge generalizing the  original $\SU(2)$ holonomies of twisted geometries and spin networks.
This allows a more complete picture of coarse-graining twisted geometries. Partitioning the network into finite connected region, we can coarse-grain each region as described in \cite{Livine:2013gna,Charles:2016xwc}. Each region is gauge-fixed to a vertex dressed with little loops. Then forgetting about the little loops, each region is coarse-grained to a vertex violating the closure constraint, with the closure defects reflecting the  curvature within the regions. Finally, we could boost each such vertex so that the boosted fluxes around the vertex satisfy once again the closure constraint. This local Lorentz transformation takes us out of the usual time gauge used in loop quantum gravity and changes the local embedding of the canonical hypersurface in the 3+1-d space-time. The coarse-graining of the links between those regions was missing up to now.
As described here, coarse-grained links themselves carry curvature excitations, which are translated into $\SL(2,\C)$ holonomies along the edges.  The final picture consists in 3d fluxes living on the edges around each vertex, a boost variable at each vertex indicated the preferred frame in which the fluxes satisfy the closure constraint, and $\SL(2,\C)$ holonomies along the links between vertices, thereby defining {\it boosted twisted geometries}. At the quantum level, these would define {\it boosted spin networks}, which seem a priori to be projected spin networks as defined in \cite{Dupuis:2010jn} but with the slight generalization that they would be evaluated on $\SL(2,\C)$ and that we would allow different $\SU(2)$ spins at the two ends of every link.

It would be enlightening to clarify the definition of those boosted spin networks and finally describe a consistent coarse-graining flow for loop quantum gravity using these structures. In particular, we would need to provide those states of geometry with a quantum gravity dynamics.
In the canonical framework, we would need to investigate how to write loop quantum gravity dynamics out of the time gauge for such projected spin network states. This would allow to explore the physics of local Lorentz transformations and clarify the covariant properties of states of geometry at the quantum level in loop quantum gravity.
In a path integral framework,  the EPRL spinfoam amplitudes \cite{Engle:2007wy,Rovelli:2010vv,Ding:2010ye,Ding:2010fw} are straightforward to adapt to this new setting with relaxed  area matching along spin network edges by projected onto different spins at the ends of every edges as suggested in \cite{Dupuis:2010jn}. It remains to see if the renormalization of spinfoam amplitudes (e.g. \cite{Riello:2014iqb,Banburski:2014cwa,Banburski:2015kmc,Chen:2016aag}) can lead to a relevant flow for the 2-point function, i.e. for the area propagator.

Finally, an interesting point in order to further understand the role $\SL(2,\C)$ holonomies in loop quantum gravity is to clarify the physical meaning in the continuum theory of the spin diffusion and spin waves that they induce along the spin network edges, especially understand if they correspond to already known phenomenology such as gravitational waves or if they describe the propagation of different types of geometrical defects.

%\begin{itemize}
%\item any link with $q$-deformation?
%\end{itemize}

%%%%%%%%%%%%%%%%%
\section*{Acknowledgement}

Plots and numerics were realized with Wolfram Mathematica 11.3.

%%%%%%%%%%%%%%%%%%
\appendix
%

%%%%%
\section{$\SL(2,\C)$ unitary representations}
\label{sl2C}
%%%%%

The  $\sl(2,\C)$ Lie algebra is 6-dimensional and we write $J_{i}$ for the  generators of the $\su(2)$ subalgebra and $K_{i}$ for the boost generators. Their commutation relations are:
\beq
&[J_{3},J_{\pm}]=\pm J_{\pm}
\,,\quad
[J_{+},J_{-}]=2J_{3}
\,,
&
\\
&
[J_{3},K_{\pm}]=\pm  K_{\pm}
\,,\quad
[J_{3},K_{3}]=0
\,,\quad
&
\nn\\
&
[K_{3},J_{\pm}]=\pm  K_{\pm}
\,,\quad
[K_{3},K_{\pm}]=\mp  J_{\pm}
\,,\quad
&
\nn\\
&
[J_{+},K_{-}]=[K_{+},J_{-}]=2  K_{3}
\,,\quad
[K_{+},K_{-}]=-2J_{3}
\,,\quad
[J_{\pm},K_{\pm}]=0
\,.
&\nn
\eeq
Irreducible representations of the $\SL(2,\C)$ Lie group are labelled by a half-integer $n\in\f\N2$  and a complex number $\rho$. We introduce the standard basis for the corresponding Hilbert space $\cR^{(n,\rho)}$ with basis states diagonalizing the $\SU(2)$ Casimir operator $\vJ^2$ and the rotation generator $J_{3}$:
\be
\cR^{(n,\rho)}
=\bigoplus_{j\ge n} \cV^j 
=\bigoplus_{j\ge n} \,\bigoplus_{-j\le m \le j} \C\,|j,m\ra\,.
\ee
The action of the $\sl(2,\C)$ generators are (see e.g. \cite{rühl1970lorentz,Alexandrov:2002br,Conrady:2010sx,naimark2014linear} and references therein):
\beq
J_{3}\,|j,m\ra
&=&
m\,|j,m\ra
\\
J_{+}\,|j,m\ra
&=&
\sqrt{(j-m)(j+m+1}\,|j,m+1\ra
\nn\\
J_{-}\,|j,m\ra
&=&
\sqrt{(j+m)(j-m+1}\,|j,m-1\ra
\nn\\
K_{3}\,|j,m\ra
&=&
\alpha_{j}\sqrt{(j-m)(j+m)}\,|j-1,m\ra
+\beta_{j}m\,|j,m\ra
+\alpha_{j+1}\sqrt{(j-m+1)(j+m+1)}\,|j+1,m\ra
\\
K_{+}\,|j,m\ra
&=&
\alpha_{j}\sqrt{(j-m)(j-m-1)}\,|j-1,m+1\ra
+\beta_{j}\sqrt{(j-m)(j+m+1}\,|j,m+1\ra
\nn\\&&
+\alpha_{j+1}\sqrt{(j+m+1)(j+m+2)}\,|j+1,m+1\ra
\nn\\
K_{-}\,|j,m\ra
&=&
-\alpha_{j}\sqrt{(j+m)(j-m-1)}\,|j-1,m-1\ra
+\beta_{j}\sqrt{(j+m)(j-m+1}\,|j,m-1\ra
\nn\\&&
-\alpha_{j+1}\sqrt{(j-m+1)(j-m+2)}\,|j+1,m-1\ra
\nn
\eeq
with the coefficients $\alpha_{j}$ and $\beta_{j}$ given in terms of the $\su(2)$ spin $j$ and the $\sl(2,\C)$ representation labels $(n,\rho)$:
\be
\alpha_{j}=\f i{2}\sqrt{\f{(j^2-n^2)(j^2+\rho^2)}{j^2(j^2-\f14)}}
\,,\qquad
\beta_{j}=\f{n\rho}{j(j+1)}
\,.
\ee
Irreducible unitary representations are classified as two series:
\begin{itemize}
\item the principal series with $\rho\in\R$ enter the Plancherel decomposition for $L^2$ functions over $\SL(2,\C)$ (with respect to the Haar measure),
\item the secondary series with $\rho\in i\R$ and bounded modulus $|\rho|<1$.
\end{itemize}
Simple representations correspond to the case $n=0$, in which case the Hilbert space $\cR^{(n,\rho)}$ carry arbitrary $\SU(2)$ spin $j\in\N$. In particular, they contain a $\SU(2)$-invariant vector and arises in the Plancherel decomposition of $L^2$ functions over the quotient space $\SL(2,\C)/\SU(2)$ \cite{DePietri:1999bx,Freidel:1999rr,Perez:2000ec,Barrett:1999qw}. They are especially relevant to the Barrett-Crane spinfoam model for a path integral formulation of quantum gravity as a quasi-topological state-sum \cite{Barrett:1997gw,Barrett:1999qw}.

%%%%%%%%%%
% BIBLIOGRAPHY

\bibliographystyle{bib-style}
\bibliography{LQG}

\end{document}